\documentclass[apj,appendixfloats,numberedappendix]{emulateapj}

\usepackage[fleqn]{amsmath}
\usepackage{wasysym}
\usepackage{rotating}
\usepackage[dvips]{color}
\usepackage{verbatim}
\usepackage{soul}
\usepackage[hypertex]{hyperref}
\usepackage{lineno}
\usepackage{longtable}
\usepackage{subfigure}

\slugcomment{Accepted for publication in ApJ}

\shorttitle{Exploring the Variability of 1633+382. Physical Properties}

\shortauthors{Algaba et al.}

\begin{document}


\title{Exploring the Variability of the Flat Spectrum Radio Source 1633+382.\\ II: Physical Properties}

\author{
Juan-Carlos Algaba$^{1,2}$, Sang-Sung Lee$^{2,3}$, Bindu Rani$^{4\dagger}$, Dae-Won Kim$^1$, Motoki Kino$^{5,6}$, Jeffrey Hodgson$^2$ , Guang-Yao Zhao$^2$, Do-Young Byun$^2$, Mark Gurwell$^7$, Sin-Cheol Kang$^{2,3}$, Jae-Young Kim$^{1,8}$, Jeong-Sook Kim$^5$, Soon-Wook Kim$^{2,3}$, Jong-Ho Park$^1$, Sascha Trippe$^1$, Kiyoaki Wajima$^2$
} 
\thanks{$^{\dagger}$NASA Postdoctoral Program (NPP) Fellow}

\affil{
$^1$Department of Physics and Astronomy, Seoul National University, 1 Gwanak-ro, Gwanak-gu, Seoul 08826, Korea\\
$^2$Korea Astronomy \& Space Science Institute, 776, Daedeokdae-ro, Yuseong-gu, Daejeon, Republic of Korea 305-348\\
$^3$Korea University of Science and Technology, 217 Gajeong-ro, Yuseong-gu, Daejeon 34113, Korea\\
$^4$NASA Goddard Space Flight Center, Greenbelt, MD 20771, USA\\
$^5$National Astronomical Observatory of Japan, 2211 Osawa, Mitaka, Tokyo 1818588, Japan\\
$^6$Kogakuin University, Academic Support Center, 2665-1 Nakano, Hachioji, Tokyo 192-0015, Japan\\
$^7$Harvard-Smithsonian Center for Astrophysics, Cambridge, MA USA\\
$^8$Max-Planck-Institut f\"ur Radioastronomie (MPIfR), Auf dem H\"ugel 69, 53121 Bonn, Germany\\
}

\begin{abstract}
The flat spectrum radio quasar 1633+382 (4C~38.41) showed a significant increase of its radio flux density during the period 2012 March -- 2015 August which correlates with $\gamma-$ray flaring activity. Multi--frequency simultaneous VLBI observations were conducted as part of the interferometric monitoring of gamma-ray bright active galactic nuclei (iMOGABA) program and supplemented with additional radio monitoring observations with the OVRO 40~m telescope, the Boston University VLBI program, and the Submillimeter Array. The epochs of the maxima for the two largest $\gamma-$ray flares coincide with the ejection of two respective new VLBI components. Analysis of the spectral energy distribution indicates a higher turnover frequency after the flaring events. The evolution of the flare in the turnover frequency--turnover flux density plane probes the adiabatic losses in agreement with the shock--in--jet model. Derived synchrotron self absorption magnetic fields, of the order of 0.1~mG, do not seem to dramatically change during the flares, and are much smaller, by a factor $10^4$, than the estimated equipartition magnetic fields, indicating that the source of the flare may be associated with a particle dominated emitting region. 
\end{abstract}


\section{Introduction}
The source 1633+382 (4C~38.41) is a flat spectrum radio quasar (FSRQ) at a redshift $z=1.813$ \citep{Hewett10}. Strong variability in its radio flux density has been observed \citep{SpanglerCotton81,Kuhr81,Seielstad85,Aller92} and superluminal motion with jet velocities up to $373\pm16~\mu$as~yr$^{-1}$ ($29.2\pm1.3~c$) has been detected \citep{Lister13}. Multi--frequency observations of the $\gamma-$ray flares observed by the \emph{Fermi}/LAT (Large Area Telescope) in 2009--2010 suggested that their origin was associated with an emerging component from the core downstream the jet at the 43~GHz VLBI core \citep{Jorstad11}. A large outburst observed in 2011 was explained geometrically due to variations of the Doppler factor owing to changes in the viewing angle \citep{Raiteri12}.

This paper is the second of a series where we study the broadband flaring activity of 1633+382 observed from 2012 to 2015. In \cite{PaperI} (hereafter, Paper~I) we studied the multi--frequency phenomenology of the light curves noting that the major $\gamma-$ray flares occurring during this period were well matched with similar enhanced activity at optical and radio frequencies. A later $\gamma-$ray flare in early 2015 was observed at X--rays and optical bands but was missing its radio frequency counterparts. For the well-associated flares, we found a significant correlation between high-energy ($\gamma-$ray, X--ray, and optical) and radio flux density variations with the former leading the later by $\sim$90 days. Using a simple model, we estimated that the distance from the high energy ($\gamma-$rays, X--rays and optical) and radio emitting regions was of the order of 40~pc during the flaring period.

The flux enhancement behaviour and phenomenology seen in 1633+382 are well understood, but the physical mechanisms and properties of the emitting regions during this period have not yet been discussed. The behaviour of the larger flux enhancements seems to be in agreement with the shock-in-jet model \citep{MarscherGear85}, where the outburst is due to a shock wave passing through the relativistic jet. Whereas it has been seen that this model may explain some of the flares historically seen in this source \citep[see e.g.][]{Jorstad11}, it seems that this may not be necessary the case for all the flaring events and, even more, alternative explanations, such as viewing angle variations \citep{Raiteri12} may also be the case.
 
In the \cite{MarscherGear85} model, electrons are accelerated at the shock front, after which they lose energy due to adiabatic expansion leading to energy stratification. In a flare, this leads to different time lags at various frequencies and steepening of the spectrum. Several sources, such as 3C~273 \citep{Turler2000,Chidiac16}; 0716+714 \citep{Rani15}; 3C~279 \citep{Lindfors06} among others have been satisfactory modelled with this description. In this model,  the magnetic field component parallel to the shock front is compressed, leading to very characteristic polarization directions transverse to the jet axis. On the other hand, if the shocked structure follows a spiral path through a helical magnetic field along the jet, a rotation of the polarization angle takes place. This has been observed for example in 1510-089 \citep{Marscher10}, 3C~279 \citep{Abdo10} or 0946+006 \citep{Itoh13}.

This prescription is however not applicable for all the cases. For example, \cite{Nagai12} found no clear correlation between $\gamma-$ray and radio light curves in 3C~84 on time scales from days to weeks, nor new components or morphology changes in the VLBI images associated with $\gamma-$ray flares. \cite{Jorstad01} found that, for 23 cases of $\gamma-$ray flares with sufficient VLBA data, only 10 of the flares (in 8 objects) fell within $1\sigma$ uncertainties of the birth epoch of a radio component. More recently, \cite{LeonTavares11} suggested that the flux relation between $\gamma-$rays and 37~GHz is positively correlated for quasars but does not exist for BL~Lacs.

In this paper we study the physical properties of the base of the jet in 1633+382  from 2012 to 2015. Investigation of the radio--morphology and the spectral energy distribution obtained with (quasi-)simultaneous multiwavelength observations, as well as the inferred magnetic fields will allow us to discuss the physical origin of the $\gamma-$ray flares that were observed during these epochs, as well as to provide a physical background and mechanisms to explain the different correlation with radio bands found in Paper~I.

The contents of this paper is organized as follows: In Section 2 we summarize the compiled and analyzed observations; in Section 3 we compile our results regarding VLBI components structure, radio spectral energy distribution and magnetic field evolution. In Section 4 we discuss the implications of these results in the context of a shock--in--jet model with a particle injection as the source of the $\gamma-$ray flares. Section 5 summarizes our conclusions.

\section{Observations and Data Analysis}
\label{dataanalysis}

In order to connect the $\gamma-$ray variability of 1633+382 with the radio flux density and morphology, we collected data from various ground-- and space--based instruments between March 2012 and August 2015 (MJD~56000 to 57250). In Table \ref{obs-summary} we summarize the various frequencies and instruments. 

\begin{table}
\begin{center}
\caption{List of Observations}
\label{obs-summary}
\begin{tabular}{ccc}
\tableline
\hline
Band &  Instrument & Frequency (Hz)\\
(1)&(2)&(3)\\
\tableline
Radio & OVRO& $1.50\times10^{10}$\\
Radio & KVN & $2.20\times10^{10}$\\
Radio & KVN/VLBA & $4.30\times10^{10}$\\
Radio & KVN & $8.60\times10^{10}$\\
Radio & KVN & $1.29\times10^{11}$\\
Radio & SMA & $2.25\times10^{11}$\\
$\gamma-$rays & {\it Fermi}--LAT & $2.42\times10^{22}-7.25\times10^{25}$\\	
	
\tableline
\end{tabular}
\end{center}
\end{table}
  
A summary is as follows: public 15~GHz monitoring data from the Owens Valley Radio Observatory (OVRO) 40~m telescope\footnote{\url{http://www.astro.caltech.edu/ovroblazars}} were used.  Simultaneous observations at 22, 43, 86 and 129~GHz were obtained with the Korean VLBI Network (KVN) under the Interferometric Monitoring of Gamma-ray Bright AGN (iMOGABA)\footnote{\url{http://radio.kasi.re.kr/sslee/}} \citep{Algaba15,Lee16}. Observations at 43~GHz were complemented with the VLBA Boston University (BU) Blazar program\footnote{\url{http://www.bu.edu/blazars/VLBAproject.html}}. Observations at 225~GHz were obtained from the Sub-millimeter array (SMA) calibrator sources list\footnote{\url{http://sma1.sma.hawaii.edu/callist/callist.html}}.  Weekly binned light curves from the {\it Fermi}-LAT (Large Area Telescope)  observed in survey mode were used for the $\gamma-$ray data\footnote{\url{https://fermi.gsfc.nasa.gov/ssc/data/access/}}. A detailed description of the observations and data reduction is given in Paper~I.

We note that, whereas iMOGABA and SMA observations are interferometric, OVRO observations at 15~GHz are single--dish, and thus may include emission from more extended regions. For a proper flux density comparison, we need to estimate how much flux density can be attributed to the innermost regions. A way to investigate this is to compare simultaneous single dish and interferometric observations. For this, we obtained archival MOJAVE and OVRO data at 15~GHz and analysed their total integrated flux densities in nearby epochs. Results are summarized in Table \ref{OVROvsMOJAVE}, where Columns 1 and 2 show the epoch and flux densities for OVRO measurements, Columns 3 and 4 show epoch and flux densities for MOJAVE observations, Column 5 shows the difference in days between the two observations, and Column 6 indicates the observed flux density difference.

\begin{table}
\begin{center}
\caption{OVRO and MOJAVE Flux Density Comparison}
\label{OVROvsMOJAVE}
\begin{tabular}{cccccc}
\tableline
\hline
OV. MJD	&	$S^{O}_{\nu}$(Jy)	&	MO. MJD	&	$S^{M}_{\nu}$(Jy)	&	$\Delta$days	&	$\Delta S_{\nu}$ \\
(1)&(2)&(3)&(4)&(5)&(6)\\
\tableline
54951	&	2.80	&	54953	&	2.89	&	-2	&	-3.0	\% \\
55007	&	2.60	&	55002	&	2.83	&	5	&	-8.2	\% \\
55062	&	2.81	&	55062	&	2.84	&	0	&	-1.3	\% \\
55191	&	3.53	&	55191	&	3.35	&	0	&	5.5	\% \\
55262	&	3.98	&	55265	&	3.92	&	-3	&	1.5	\% \\
55458	&	3.63	&	55468	&	3.67	&	-10	&	-1.1	\% \\
55497	&	3.62	&	55494	&	3.52	&	3	&	2.9	\% \\
55613	&	3.51	&	55612	&	3.52	&	1	&	-0.3	\% \\
55706	&	3.40	&	55702	&	3.44	&	4	&	-1.3	\% \\
55738	&	3.33	&	55736	&	3.28	&	2	&	1.4	\% \\
55789	&	3.27	&	55788	&	3.23	&	1	&	1.3	\% \\
55929	&	2.85	&	55928	&	2.92	&	1	&	-2.2	\% \\
56021	&	2.51	&	56013	&	2.56	&	8	&	-1.9	\% \\
56244	&	4.19	&	56242	&	3.95	&	2	&	5.9	\% \\
56260	&	4.19	&	56259	&	4.00	&	1	&	4.7	\% \\
56334	&	3.88	&	56333	&	3.75	&	1	&	3.4	\% \\
56506	&	4.50	&	56503	&	4.33	&	3	&	3.8	\% \\
56714	&	3.91	&	56715	&	4.13	&	-1	&	-5.4	\% \\
57146	&	2.35	&	57150	&	2.36	&	-4	&	-0.1	\% \\
57402	&	3.13	&	57403	&	3.12	&	-1	&	0.2	\% \\
\tableline
\end{tabular}
\end{center}
\end{table}

Time difference is as small as a few days and, in some cases, observations were simultaneous, so little variation can be ascribed to source variability. Flux density difference is sometimes positive, indicating larger flux density observed from single dish measurements, and other times negative, indicating that the flux density appears to be larger on VLBI measurements. Dispersion on the flux density difference is 3.6\%, larger than typical SMA errors of $\sim1$\% but smaller than the typical VLBA errors of $\sim5$\%. This indicates that most of the single--dish flux density arises from the VLBI regions or that the difference is dominated by the VLBI amplitude calibration uncertainties. We thus conclude that the results presented here are not sensitive to a combined usage of single dish and VLBI flux densities.

In order to study the innermost structure of 1633+382, we analyzed the high--resolution radio--images available with the interferometric instruments (i.e., KVN and VLBI). Under the iMOGABA view, with a typical resolution up to $\lesssim$1~mas at 129~GHz, 1633+382 still appears as a point-like unresolved source, and even circular Gaussian model-fitting is not able to reveal reliable extended structure above the resolution and sensitivity limits of the KVN iMOGABA images for this source. We thus used the 43~GHz VLBA BU Blazar program images, with typical resolutions of  $\sim0.2$~mas and dynamic ranges DR$>10^3$, to extract the information about the high--resolution structure of this source.

\begin{figure}
\includegraphics[scale=0.5]{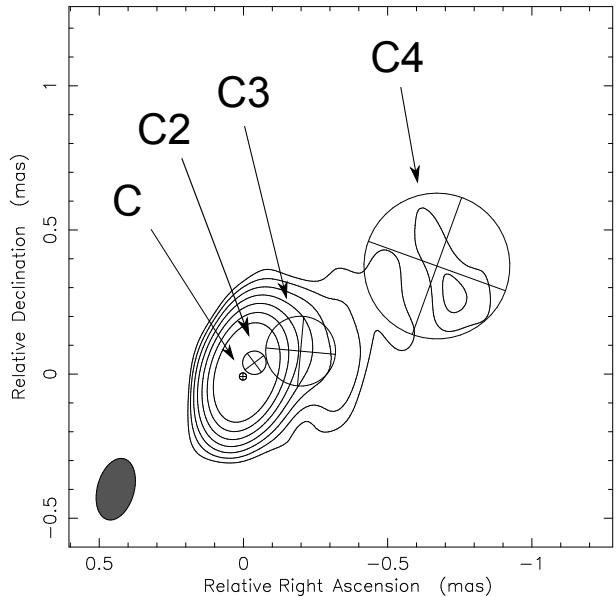}
\caption{Example of model-fitted map obtained with the 43~GHz BU VLBA Blazar program data, with the different components described in the text marked.  Lowest contours correspond to 3 times the level noise of the map and increase in level of 0.015 Jy/beam $\times$(1, 2, 4, 8, 16, 32, 64). The map peak flux is 4.8 Jy/beam. The grey ellipse in the bottom left indicates the beam size. Circles with crosses show the fitted components, their position and size.}
\label{SampleModelFittedMap}
\end{figure}

\cite{Jorstad17} presents a modelling of total intensity images for each epoch by circular Gaussian components that best fits the visibility data by using the \textsc{modelfit} task in the Caltech \emph{Difmap} package \citep{Shepherd97}. However, they present information of epochs only up to MJD~56308, whereas here we discuss data that extends in time more than two additional years. We thus independently model--fitted the various components of their VLBA data. We fitted a number of circular Gaussians taking into account the residual map side lobes, noise levels, and reduced chi-square, as well as compatibility with nearby epochs, including two cases where we can compare with the same-epoch BU model-fitted map. An example of the model--fitted image is shown in Figure \ref{SampleModelFittedMap} for the map observed in 2013 November 18th, which is directly comparable to the CLEANed image obtained by the BU blazar program\footnote{\url{https://www.bu.edu/blazars/VLBA\_GLAST/1633/1633nov13\_map.jpg}}. In addition to the core, we were able to fit various other components that we will hereafter identify as C2, C3, C4 and C5. We compared our model-fitted parameters (distance to the core, position angle, size and flux) with these provided in \cite{Jorstad17} for these epochs where information is available. We concluded that, although some small differences are found, our models agree well within the given uncertainties.

We took into account the resolution limits as follows. The minimum resolvable size of a component in a general VLBI image is given by \citep{Lobanov05}:
\begin{equation}
d_{\rm min}=2^{1+\beta/2}{\left[{\frac{ab\ln2}{\pi}\ln\left({\frac{SNR}{SNR-1}}\right)}\right]}^{1/2}, 
\label{resolutionlimits}
\end{equation}
where {\it a} and {\it b} are the axes of the restoring beam of observations, {\it SNR} is the signal-to-noise ratio of the jet component, and $\beta$ is a weighting function of imaging, which is 0 for natural weighting or 2 for uniform weighting. If $d<d_{\rm min}$ for a component, then the component is considered to be unresolved. Otherwise, an estimation of its size and distance errors $\sigma_d$, $\sigma_r$ can be given by 
$\sigma_d\sim d/\sqrt{\rm DR}$ and $\sigma_r\sim(1/2)\sigma_d$, for DR$\gg1$ \citep[see e.g.][]{Fomalont99,Lee08}.

For the kinematic analysis, we took the core position as the reference point for the rest of the components. We note however that the absolute position of the core may not be constant over time, due to changes in the opacity, instabilities in the upstream regions of the jet or other effects. Inaccuracies rising from this will be however very small in general and will cause negligible effects in our study. For example, 
investigating the time variation of the core position offsets of 3C~454.3 shown in \cite{Mohan15}, we find a standard deviation of 44, 30, 16 and 13~$\mu$as for 4.8, 8, 14.5 and 22~GHz respectively. Extrapolation to 43~GHz indicates that a maximum deviation of $7$~$\mu$as is expected for the BU VLBA data, which is smaller than our error estimation. Additionally, the analysis in \cite{Algaba12} and  Paper~I seem to indicate that core shift effects are small in this source, and thus epoch--to--epoch changes in the core position due to core--shift will be negligible. 

\section{Results}

\subsection{VLBA component structure}

\begin{figure}
\includegraphics[scale=0.46,trim={0cm 0cm 0cm 0cm},clip]{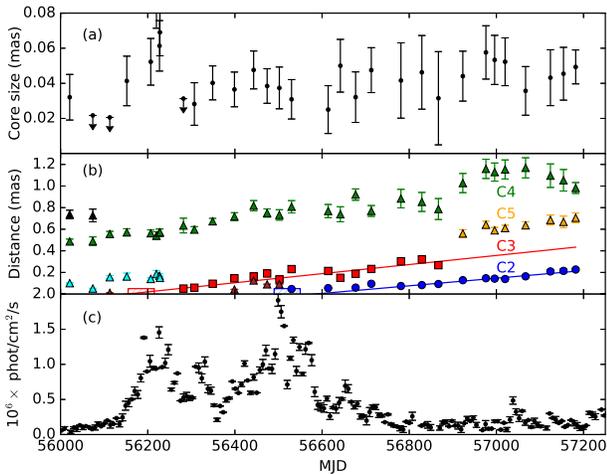}
\caption{(a) Boston University VLBI 43~GHz core size. (b) BU VLBI 43~GHz components distance from the core (each component is identified with a different color and symbol: C2 blue circles; C3 red squares; C4 green triangles). Straight line indicates the linear fit to the components C2 and C3, and hatched boxes (with arbitrary height) indicate the errors in the ejection epoch. (c) $\gamma-$rays light curve.}
\label{fig_components}
\end{figure}

In Table \ref{table_components} we summarize the results of the model fitting for the 43~GHz VLBA BU Blazar program images. Columns 1 and 2 show the epoch in Gregorian and modified Julian dates, respectively. Column 3 indicates the identification the component can be associated with (C= core, C2, C3, C4 and C5 for the long-lived clearly identified components, B4 for the relatively short-lived component that we can cross-identify with BU data, and CX for other short-lived components). Columns 4, 5 and 6 indicate the model-fitted flux, radial distance, position angle and size of the observed components. In Figure \ref{fig_components} we plot the time evolution of the core size and components distances. The $\gamma-$ray light curve is also shown for comparison and reference.

\begin{longtable*}{ccccccc}
\caption{\\ \textsc{Model-fitted Components Properties}}
\label{table_components}
\\ \hline \\ \hline
Epoch & MJD & Component & Flux (Jy) & Distance (mas) & P.A. (deg) & Size (mas) \\
(1)&(2)&(3)&(4)&(5)&(6)&(7)\\ \hline 
\endfirsthead

\multicolumn{7}{c}%
{{ \tablename\ \thetable{} -- continued from previous page}} \\ \hline \\ \hline
Epoch & MJD & Component & Flux (Jy) & Distance (mas) & P.A. (deg) & Size (mas) \\
(1)&(2)&(3)&(4)&(5)&(6)&(7)\\ \hline
\endhead

\\ \hline \multicolumn{7}{c}{{Continued on next page}} \\ 
\endfoot

\\ \hline \\ \hline
\endlastfoot
2012 Apr 03	&	56021	&	C	&$	1.64	\pm	0.05	$&$	...			$&$	...			$&$	0.03	\pm	0.01	$\\
2012 Apr 03	&	56021	&	B4	&$	0.11	\pm	0.01	$&$	0.10	\pm	0.01	$&$	246	\pm	4	$&$	0.11	\pm	0.01	$\\
2012 Apr 03	&	56021	&	C4	&$	0.20	\pm	0.02	$&$	0.49	\pm	0.02	$&$	-57	\pm	3	$&$	0.27	\pm	0.04	$\\
2012 Apr 03	&	56021	&	CX	&$	0.14	\pm	0.02	$&$	0.74	\pm	0.04	$&$	-52	\pm	3	$&$	0.53	\pm	0.08	$\\
2012 May 27	&	56074	&	C	&$	1.65	\pm	0.09	$&$	...			$&$	...			$&$	<		0.02	$\\
2012 May 27	&	56074	&	B4	&$	0.47	\pm	0.05	$&$	0.06	\pm	0.00	$&$	237	\pm	3	$&$	0.07	\pm	0.01	$\\
2012 May 27	&	56074	&	C4	&$	0.17	\pm	0.03	$&$	0.50	\pm	0.04	$&$	-59	\pm	5	$&$	0.24	\pm	0.08	$\\
2012 May 27	&	56074	&	CX	&$	0.18	\pm	0.03	$&$	0.73	\pm	0.06	$&$	-53	\pm	5	$&$	0.52	\pm	0.12	$\\
2012 Jul 05	&	56113	&	C	&$	1.15	\pm	0.06	$&$	...			$&$	...			$&$	<		0.02	$\\
2012 Jul 05	&	56113	&	CX	&$	1.89	\pm	0.07	$&$	0.03	\pm	0.00	$&$	125	\pm	1	$&$	0.05	\pm	0.00	$\\
2012 Jul 05	&	56113	&	B4	&$	0.08	\pm	0.02	$&$	0.17	\pm	0.01	$&$	253	\pm	5	$&$	0.16	\pm	0.03	$\\
2012 Jul 05	&	56113	&	C4	&$	0.37	\pm	0.03	$&$	0.58	\pm	0.02	$&$	-59	\pm	3	$&$	0.40	\pm	0.05	$\\
2012 Aug 13	&	56152	&	C	&$	3.14	\pm	0.11	$&$	...			$&$	...			$&$	0.04	\pm	0.01	$\\
2012 Aug 13	&	56152	&	B4	&$	0.03	\pm	0.01	$&$	0.16	\pm	0.03	$&$	306	\pm	11	$&$	<		0.06	$\\
2012 Aug 13	&	56152	&	C4	&$	0.28	\pm	0.03	$&$	0.57	\pm	0.03	$&$	300	\pm	3	$&$	0.41	\pm	0.07	$\\
2012 Oct 07	&	56207	&	C	&$	3.84	\pm	0.12	$&$	...			$&$	...			$&$	0.05	\pm	0.01	$\\
2012 Oct 07	&	56207	&	B4	&$	0.13	\pm	0.02	$&$	0.13	\pm	0.01	$&$	-105	\pm	5	$&$	<		0.02	$\\
2012 Oct 07	&	56207	&	C4	&$	0.33	\pm	0.04	$&$	0.56	\pm	0.03	$&$	-60	\pm	3	$&$	0.49	\pm	0.06	$\\
2012 Oct 19	&	56220	&	C	&$	3.73	\pm	0.16	$&$	...			$&$	...			$&$	0.09	\pm	0.02	$\\
2012 Oct 19	&	56220	&	B4	&$	0.06	\pm	0.02	$&$	0.18	\pm	0.03	$&$	-108	\pm	10	$&$	<		0.07	$\\
2012 Oct 19	&	56220	&	C4	&$	0.38	\pm	0.05	$&$	0.52	\pm	0.04	$&$	-58	\pm	4	$&$	0.59	\pm	0.07	$\\
2012 Oct 27	&	56227	&	C	&$	3.49	\pm	0.12	$&$	...			$&$	...			$&$	0.06	\pm	0.01	$\\
2012 Oct 27	&	56227	&	B4	&$	0.08	\pm	0.02	$&$	0.17	\pm	0.02	$&$	-110	\pm	7	$&$	<		0.04	$\\
2012 Oct 27	&	56227	&	C4	&$	0.31	\pm	0.04	$&$	0.56	\pm	0.03	$&$	-58	\pm	3	$&$	0.47	\pm	0.07	$\\
2012 Oct 28	&	56228	&	C	&$	3.75	\pm	0.12	$&$	...			$&$	...			$&$	0.07	\pm	0.01	$\\
2012 Oct 28	&	56228	&	B4	&$	0.08	\pm	0.02	$&$	0.15	\pm	0.02	$&$	266	\pm	6	$&$	<		0.03	$\\
2012 Oct 28	&	56228	&	C4	&$	0.31	\pm	0.04	$&$	0.57	\pm	0.03	$&$	-58	\pm	3	$&$	0.46	\pm	0.07	$\\
2012 Dec 21	&	56282	&	C	&$	2.05	\pm	0.16	$&$	...			$&$	...			$&$	<		0.03	$\\
2012 Dec 21	&	56282	&	C3	&$	0.99	\pm	0.11	$&$	0.06	\pm	0.00	$&$	295	\pm	3	$&$	0.13	\pm	0.01	$\\
2012 Dec 21	&	56282	&	C4	&$	0.27	\pm	0.06	$&$	0.64	\pm	0.07	$&$	304	\pm	6	$&$	0.43	\pm	0.14	$\\
2013 Jan 15	&	56307	&	C	&$	2.35	\pm	0.07	$&$	...			$&$	...			$&$	0.03	\pm	0.01	$\\
2013 Jan 15	&	56307	&	C3	&$	0.86	\pm	0.04	$&$	0.07	\pm	0.00	$&$	287	\pm	1	$&$	0.14	\pm	0.00	$\\
2013 Jan 15	&	56307	&	C4	&$	0.23	\pm	0.02	$&$	0.61	\pm	0.03	$&$	301	\pm	3	$&$	0.46	\pm	0.06	$\\
2013 Feb 26	&	56349	&	C	&$	3.59	\pm	0.08	$&$	...			$&$	...			$&$	0.04	\pm	0.01	$\\
2013 Feb 26	&	56349	&	C3	&$	0.87	\pm	0.04	$&$	0.09	\pm	0.00	$&$	-66	\pm	1	$&$	0.16	\pm	0.00	$\\
2013 Feb 26	&	56349	&	C4	&$	0.29	\pm	0.02	$&$	0.66	\pm	0.03	$&$	-58	\pm	2	$&$	0.52	\pm	0.06	$\\
2013 Apr 17	&	56399	&	C	&$	4.28	\pm	0.10	$&$	...			$&$	...			$&$	0.04	\pm	0.01	$\\
2013 Apr 17	&	56399	&	CX	&$	0.70	\pm	0.04	$&$	0.04	\pm	0.00	$&$	-46	\pm	2	$&$	0.15	\pm	0.00	$\\
2013 Apr 17	&	56399	&	C3	&$	0.33	\pm	0.03	$&$	0.15	\pm	0.01	$&$	-68	\pm	2	$&$	0.25	\pm	0.01	$\\
2013 Apr 17	&	56399	&	C4	&$	0.25	\pm	0.03	$&$	0.72	\pm	0.04	$&$	-57	\pm	3	$&$	0.53	\pm	0.07	$\\
2013 May 31	&	56443	&	C	&$	4.37	\pm	0.11	$&$	...			$&$	...			$&$	0.05	\pm	0.01	$\\
2013 May 31	&	56443	&	CX	&$	0.29	\pm	0.03	$&$	0.12	\pm	0.01	$&$	-42	\pm	3	$&$	0.08	\pm	0.01	$\\
2013 May 31	&	56443	&	C3	&$	0.45	\pm	0.04	$&$	0.16	\pm	0.01	$&$	-68	\pm	2	$&$	0.28	\pm	0.01	$\\
2013 May 31	&	56443	&	C4	&$	0.22	\pm	0.03	$&$	0.82	\pm	0.05	$&$	-57	\pm	3	$&$	0.50	\pm	0.10	$\\
2013 Jul 01	&	56474	&	C	&$	3.50	\pm	0.08	$&$	...			$&$	...			$&$	0.04	\pm	0.01	$\\
2013 Jul 01	&	56474	&	CX	&$	0.59	\pm	0.03	$&$	0.09	\pm	0.00	$&$	-47	\pm	2	$&$	0.13	\pm	0.01	$\\
2013 Jul 01	&	56474	&	C3	&$	0.21	\pm	0.02	$&$	0.19	\pm	0.01	$&$	-73	\pm	3	$&$	0.25	\pm	0.02	$\\
2013 Jul 01	&	56474	&	C4	&$	0.20	\pm	0.02	$&$	0.75	\pm	0.04	$&$	-61	\pm	3	$&$	0.56	\pm	0.08	$\\
2013 Jul 29	&	56502	&	C	&$	4.49	\pm	0.13	$&$	...			$&$	...			$&$	0.04	\pm	0.01	$\\
2013 Jul 29	&	56502	&	CX	&$	0.95	\pm	0.06	$&$	0.09	\pm	0.00	$&$	-50	\pm	2	$&$	0.19	\pm	0.01	$\\
2013 Jul 29	&	56502	&	C3	&$	0.53	\pm	0.04	$&$	0.13	\pm	0.01	$&$	-64	\pm	2	$&$	0.18	\pm	0.01	$\\
2013 Jul 29	&	56502	&	C4	&$	0.22	\pm	0.03	$&$	0.73	\pm	0.05	$&$	-61	\pm	4	$&$	0.52	\pm	0.10	$\\
2013 Aug 26	&	56530	&	C	&$	4.18	\pm	0.11	$&$	...			$&$	...			$&$	0.03	\pm	0.01	$\\
2013 Aug 26	&	56530	&	C2	&$	1.01	\pm	0.06	$&$	0.05	\pm	0.00	$&$	299	\pm	2	$&$	0.15	\pm	0.00	$\\
2013 Aug 26	&	56530	&	C3	&$	0.18	\pm	0.02	$&$	0.23	\pm	0.02	$&$	285	\pm	4	$&$	0.24	\pm	0.03	$\\
2013 Aug 26	&	56530	&	C4	&$	0.16	\pm	0.02	$&$	0.81	\pm	0.06	$&$	300	\pm	4	$&$	0.52	\pm	0.11	$\\
2013 Nov 18	&	56614	&	C	&$	3.80	\pm	0.13	$&$	...			$&$	...			$&$	0.03	\pm	0.01	$\\
2013 Nov 18	&	56614	&	C2	&$	1.59	\pm	0.08	$&$	0.06	\pm	0.00	$&$	311	\pm	1	$&$	0.08	\pm	0.00	$\\
2013 Nov 18	&	56614	&	C3	&$	0.33	\pm	0.04	$&$	0.22	\pm	0.01	$&$	290	\pm	3	$&$	0.24	\pm	0.02	$\\
2013 Nov 18	&	56614	&	C4	&$	0.17	\pm	0.03	$&$	0.77	\pm	0.06	$&$	299	\pm	5	$&$	0.51	\pm	0.12	$\\
2013 Dec 16	&	56642	&	C	&$	4.25	\pm	0.16	$&$	...			$&$	...			$&$	0.05	\pm	0.02	$\\
2013 Dec 16	&	56642	&	C3	&$	0.52	\pm	0.05	$&$	0.14	\pm	0.01	$&$	-57	\pm	3	$&$	0.22	\pm	0.02	$\\
2013 Dec 16	&	56642	&	C4	&$	0.17	\pm	0.03	$&$	0.73	\pm	0.07	$&$	-59	\pm	5	$&$	0.56	\pm	0.14	$\\
2014 Jan 20	&	56677	&	C	&$	2.68	\pm	0.09	$&$	...			$&$	...			$&$	0.03	\pm	0.01	$\\
2014 Jan 20	&	56677	&	C2	&$	2.04	\pm	0.08	$&$	0.07	\pm	0.00	$&$	-37	\pm	1	$&$	0.07	\pm	0.00	$\\
2014 Jan 20	&	56677	&	C3	&$	0.42	\pm	0.04	$&$	0.19	\pm	0.01	$&$	-68	\pm	3	$&$	0.29	\pm	0.02	$\\
2014 Jan 20	&	56677	&	C4	&$	0.25	\pm	0.03	$&$	0.92	\pm	0.05	$&$	-54	\pm	3	$&$	0.83	\pm	0.11	$\\
2014 Feb 25	&	56713	&	C	&$	2.53	\pm	0.08	$&$	...			$&$	...			$&$	0.05	\pm	0.01	$\\
2014 Feb 25	&	56713	&	C2	&$	1.33	\pm	0.06	$&$	0.08	\pm	0.00	$&$	-23	\pm	1	$&$	0.07	\pm	0.00	$\\
2014 Feb 25	&	56713	&	C3	&$	0.38	\pm	0.03	$&$	0.22	\pm	0.01	$&$	-56	\pm	2	$&$	0.26	\pm	0.02	$\\
2014 Feb 25	&	56713	&	C4	&$	0.15	\pm	0.02	$&$	0.76	\pm	0.05	$&$	-58	\pm	4	$&$	0.46	\pm	0.10	$\\
2014 Feb 25	&	56713	&	CX	&$	0.13	\pm	0.02	$&$	1.45	\pm	0.10	$&$	-52	\pm	4	$&$	0.79	\pm	0.20	$\\
2014 May 04	&	56781	&	C	&$	1.96	\pm	0.10	$&$	...			$&$	...			$&$	0.04	\pm	0.02	$\\
2014 May 04	&	56781	&	C2	&$	0.52	\pm	0.05	$&$	0.08	\pm	0.00	$&$	306	\pm	3	$&$	0.07	\pm	0.01	$\\
2014 May 04	&	56781	&	C3	&$	0.21	\pm	0.03	$&$	0.31	\pm	0.02	$&$	289	\pm	5	$&$	0.43	\pm	0.05	$\\
2014 May 04	&	56781	&	C4	&$	0.15	\pm	0.03	$&$	0.89	\pm	0.09	$&$	303	\pm	6	$&$	0.70	\pm	0.17	$\\
2014 Jun 21	&	56829	&	C	&$	1.84	\pm	0.09	$&$	...			$&$	...			$&$	0.05	\pm	0.02	$\\
2014 Jun 21	&	56829	&	C2	&$	0.51	\pm	0.05	$&$	0.08	\pm	0.00	$&$	-66	\pm	3	$&$	0.08	\pm	0.01	$\\
2014 Jun 21	&	56829	&	C3	&$	0.13	\pm	0.03	$&$	0.32	\pm	0.03	$&$	-66	\pm	5	$&$	0.34	\pm	0.06	$\\
2014 Jun 21	&	56829	&	C4	&$	0.12	\pm	0.02	$&$	0.85	\pm	0.09	$&$	-60	\pm	6	$&$	0.53	\pm	0.18	$\\
2014 Jul 29	&	56867	&	C	&$	1.68	\pm	0.11	$&$	...			$&$	...			$&$	0.03	\pm	0.03	$\\
2014 Jul 29	&	56867	&	C2	&$	0.45	\pm	0.06	$&$	0.10	\pm	0.01	$&$	-45	\pm	4	$&$	0.06	\pm	0.01	$\\
2014 Jul 29	&	56867	&	C3	&$	0.15	\pm	0.03	$&$	0.27	\pm	0.03	$&$	-70	\pm	6	$&$	0.31	\pm	0.06	$\\
2014 Jul 29	&	56867	&	C4	&$	0.12	\pm	0.03	$&$	0.79	\pm	0.10	$&$	-59	\pm	7	$&$	0.47	\pm	0.19	$\\
2014 Sep 23	&	56923	&	C	&$	1.50	\pm	0.05	$&$	...			$&$	...			$&$	0.04	\pm	0.01	$\\
2014 Sep 23	&	56923	&	C2	&$	0.19	\pm	0.02	$&$	0.12	\pm	0.01	$&$	-64	\pm	3	$&$	0.05	\pm	0.01	$\\
2014 Sep 23	&	56923	&	C5	&$	0.11	\pm	0.01	$&$	0.56	\pm	0.04	$&$	-69	\pm	4	$&$	0.35	\pm	0.07	$\\
2014 Sep 23	&	56923	&	C4	&$	0.06	\pm	0.01	$&$	1.02	\pm	0.09	$&$	-56	\pm	5	$&$	0.59	\pm	0.18	$\\
2014 Nov 15	&	56976	&	C	&$	1.01	\pm	0.04	$&$	...			$&$	...			$&$	0.06	\pm	0.02	$\\
2014 Nov 15	&	56976	&	C2	&$	0.22	\pm	0.02	$&$	0.13	\pm	0.01	$&$	-71	\pm	2	$&$	0.17	\pm	0.01	$\\
2014 Nov 15	&	56976	&	C5	&$	0.13	\pm	0.01	$&$	0.63	\pm	0.03	$&$	-68	\pm	3	$&$	0.44	\pm	0.07	$\\
2014 Nov 15	&	56976	&	C4	&$	0.06	\pm	0.01	$&$	1.15	\pm	0.09	$&$	-52	\pm	4	$&$	0.71	\pm	0.18	$\\
2014 Dec 05	&	56996	&	C	&$	0.99	\pm	0.03	$&$	...			$&$	...			$&$	0.05	\pm	0.01	$\\
2014 Dec 05	&	56996	&	C2	&$	0.16	\pm	0.01	$&$	0.15	\pm	0.01	$&$	299	\pm	2	$&$	0.11	\pm	0.01	$\\
2014 Dec 05	&	56996	&	C5	&$	0.13	\pm	0.01	$&$	0.60	\pm	0.03	$&$	294	\pm	3	$&$	0.43	\pm	0.05	$\\
2014 Dec 05	&	56996	&	C4	&$	0.05	\pm	0.01	$&$	1.14	\pm	0.08	$&$	309	\pm	4	$&$	0.63	\pm	0.17	$\\
2014 Dec 29	&	57020	&	C	&$	0.96	\pm	0.03	$&$	...			$&$	...			$&$	0.05	\pm	0.01	$\\
2014 Dec 29	&	57020	&	C2	&$	0.16	\pm	0.01	$&$	0.14	\pm	0.01	$&$	295	\pm	2	$&$	0.13	\pm	0.01	$\\
2014 Dec 29	&	57020	&	C5	&$	0.14	\pm	0.01	$&$	0.62	\pm	0.03	$&$	294	\pm	2	$&$	0.44	\pm	0.05	$\\
2014 Dec 29	&	57020	&	C4	&$	0.05	\pm	0.01	$&$	1.16	\pm	0.09	$&$	310	\pm	4	$&$	0.60	\pm	0.18	$\\
2015 Feb 14	&	57067	&	C	&$	1.18	\pm	0.04	$&$	...			$&$	...			$&$	0.04	\pm	0.01	$\\
2015 Feb 14	&	57067	&	C2	&$	0.14	\pm	0.01	$&$	0.17	\pm	0.01	$&$	289	\pm	3	$&$	0.19	\pm	0.02	$\\
2015 Feb 14	&	57067	&	C5	&$	0.13	\pm	0.01	$&$	0.64	\pm	0.03	$&$	292	\pm	3	$&$	0.42	\pm	0.07	$\\
2015 Feb 14	&	57067	&	C4	&$	0.05	\pm	0.01	$&$	1.17	\pm	0.09	$&$	308	\pm	5	$&$	0.71	\pm	0.19	$\\
2015 Apr 12	&	57124	&	C	&$	1.51	\pm	0.06	$&$	...			$&$	...			$&$	0.04	\pm	0.02	$\\
2015 Apr 12	&	57124	&	C2	&$	0.13	\pm	0.02	$&$	0.21	\pm	0.01	$&$	-69	\pm	4	$&$	0.21	\pm	0.03	$\\
2015 Apr 12	&	57124	&	C5	&$	0.13	\pm	0.02	$&$	0.68	\pm	0.05	$&$	-70	\pm	4	$&$	0.40	\pm	0.09	$\\
2015 Apr 12	&	57124	&	C4	&$	0.06	\pm	0.01	$&$	1.10	\pm	0.11	$&$	-51	\pm	6	$&$	0.60	\pm	0.22	$\\
2015 May 12	&	57154	&	C	&$	1.69	\pm	0.06	$&$	...			$&$	...			$&$	0.05	\pm	0.02	$\\
2015 May 12	&	57154	&	C2	&$	0.09	\pm	0.01	$&$	0.22	\pm	0.02	$&$	296	\pm	5	$&$	0.20	\pm	0.04	$\\
2015 May 12	&	57154	&	C5	&$	0.09	\pm	0.01	$&$	0.67	\pm	0.05	$&$	288	\pm	5	$&$	0.30	\pm	0.11	$\\
2015 May 12	&	57154	&	C4	&$	0.06	\pm	0.01	$&$	1.06	\pm	0.10	$&$	307	\pm	5	$&$	0.59	\pm	0.20	$\\
2015 Jun 09	&	57182	&	C	&$	1.91	\pm	0.04	$&$	...			$&$	...			$&$	0.05	\pm	0.01	$\\
2015 Jun 09	&	57182	&	C2	&$	0.12	\pm	0.01	$&$	0.23	\pm	0.01	$&$	297	\pm	3	$&$	0.24	\pm	0.02	$\\
2015 Jun 09	&	57182	&	C5	&$	0.06	\pm	0.01	$&$	0.71	\pm	0.05	$&$	288	\pm	4	$&$	0.22	\pm	0.09	$\\
2015 Jun 09	&	57182	&	C4	&$	0.10	\pm	0.01	$&$	0.98	\pm	0.05	$&$	305	\pm	3	$&$	0.63	\pm	0.10	$\\
\end{longtable*}

Various things are immediately noticeable from inspection of this figure. First, all components appear to be moving away from the core. If we perform a simple linear fit of the distance from the core for components C2 and C3, we find that C2 appears to be ejected from the core at MJD$=56585\pm30$ with speed $130\pm10$~$\mu$as~yr$^{-1}$ ($10.2\pm0.8$~c); whereas C3 seems to be have been ejected from the core at MJD$=56153\pm30$ with speed $150\pm20$~$\mu$as~yr$^{-1}$ ($11.7\pm1.6$~c). These component ejection epochs are very close and compatible with the dates corresponding to the maxima of the $\gamma-$ray flares, and may pinpoint the structural origin of these flares, as discussed below.

If we assume that the emitted components move towards the observer with a viewing angle of 2.5\degr \citep{Hovatta09,Liu10}, the calculated apparent speeds lead to Lorentz factors $\Gamma\sim12-14$ and Doppler factors $\delta\sim19-21$, respectively. The estimated values are in agreement with the value for $\delta_{var}\sim21.5$ found by \cite{Hovatta09}. 

Second, the apparent deconvolved size of the core seems to be roughly constant with time, although little variations can be appreciated. Some epochs at the beginning of our analysis show an unresolved core, although this could be due to blending effects. There seems not to be an obvious behaviour related with the flux enhancement or component ejection. The median core size found is about 0.04~mas.

\subsection{Radio Spectral Energy Distribution}

We investigate here the radio spectral energy distribution (SED) of the radio flux densities. Once the possible differences in flux densities due to interferometric versus single dish observations have been taken into consideration (see Section \ref{dataanalysis}), the methodology to obtain the SED is as follows. First, we concentrated in these epochs for which simultaneous data for iMOGABA are provided. Then we searched for the closest epochs for OVRO at 15~GHz and SMA at 225~GHz (see Paper~I). If these epochs were within two weeks (14 days) of the iMOGABA data, we then considered them to be quasi--simultaneous and included its flux density to consider the SED.  We note that a maximum of six points could be obtained, but in some epochs we could not include quasi-simultaneous data from  SMA 225~GHz or iMOGABA 129~GHz. We restricted our study for these epochs where at least 5 data points could be obtained. Likewise, epochs where iMOGABA was under maintenance, or no observations where made, but data with OVRO 15~GHz, BU 43~GHz and SMA 225~GHz were ready, where not considered as they would consist only of 3 data points. In this way we obtained SEDs for 17 different epochs.

\begin{figure*}
\includegraphics[height=4cm, width=6cm,trim=0cm 0cm 0cm 0cm,clip=true]{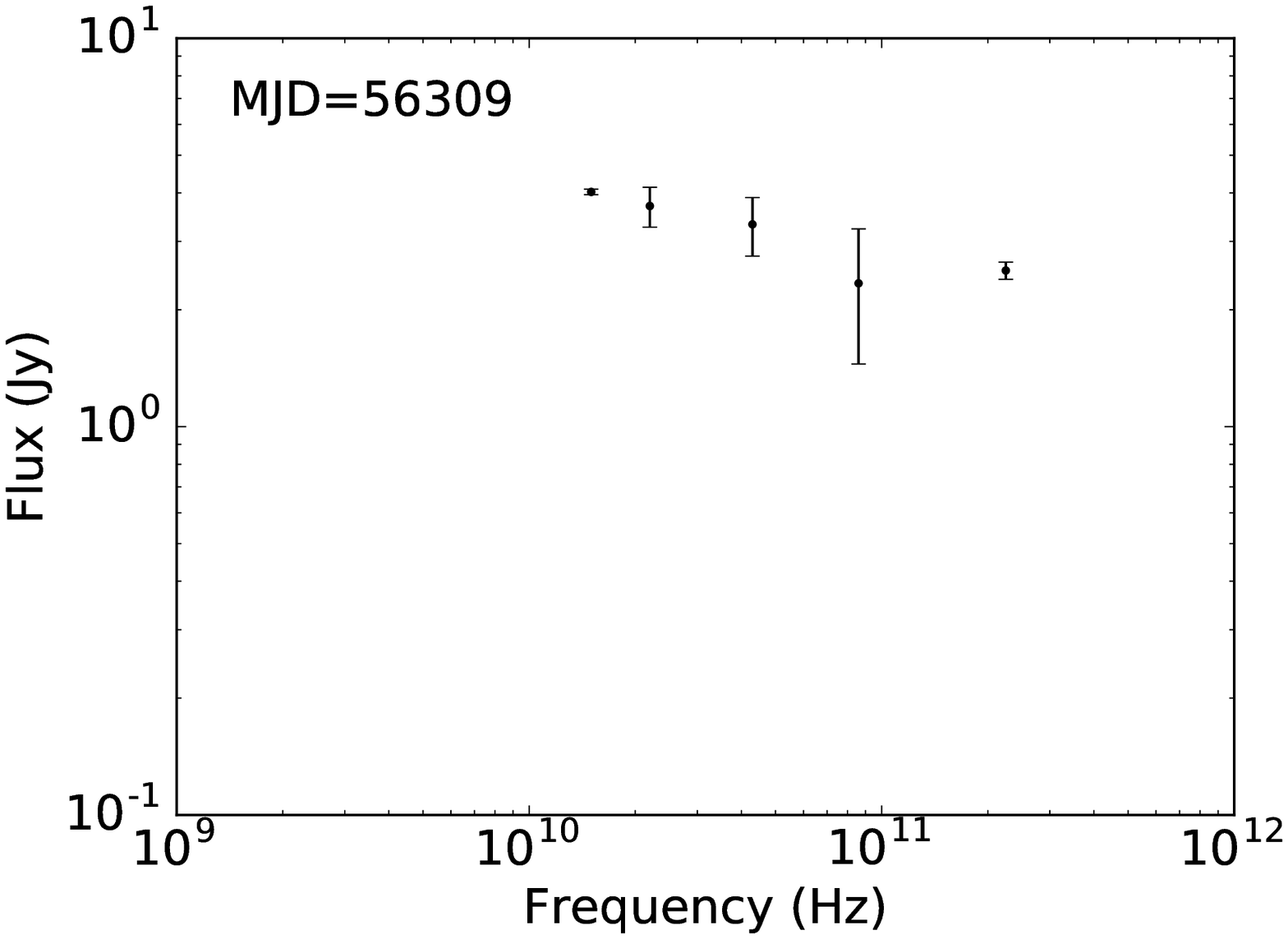}
\includegraphics[height=4cm, width=6cm,trim=0cm 0cm 0cm 0cm,clip=true]{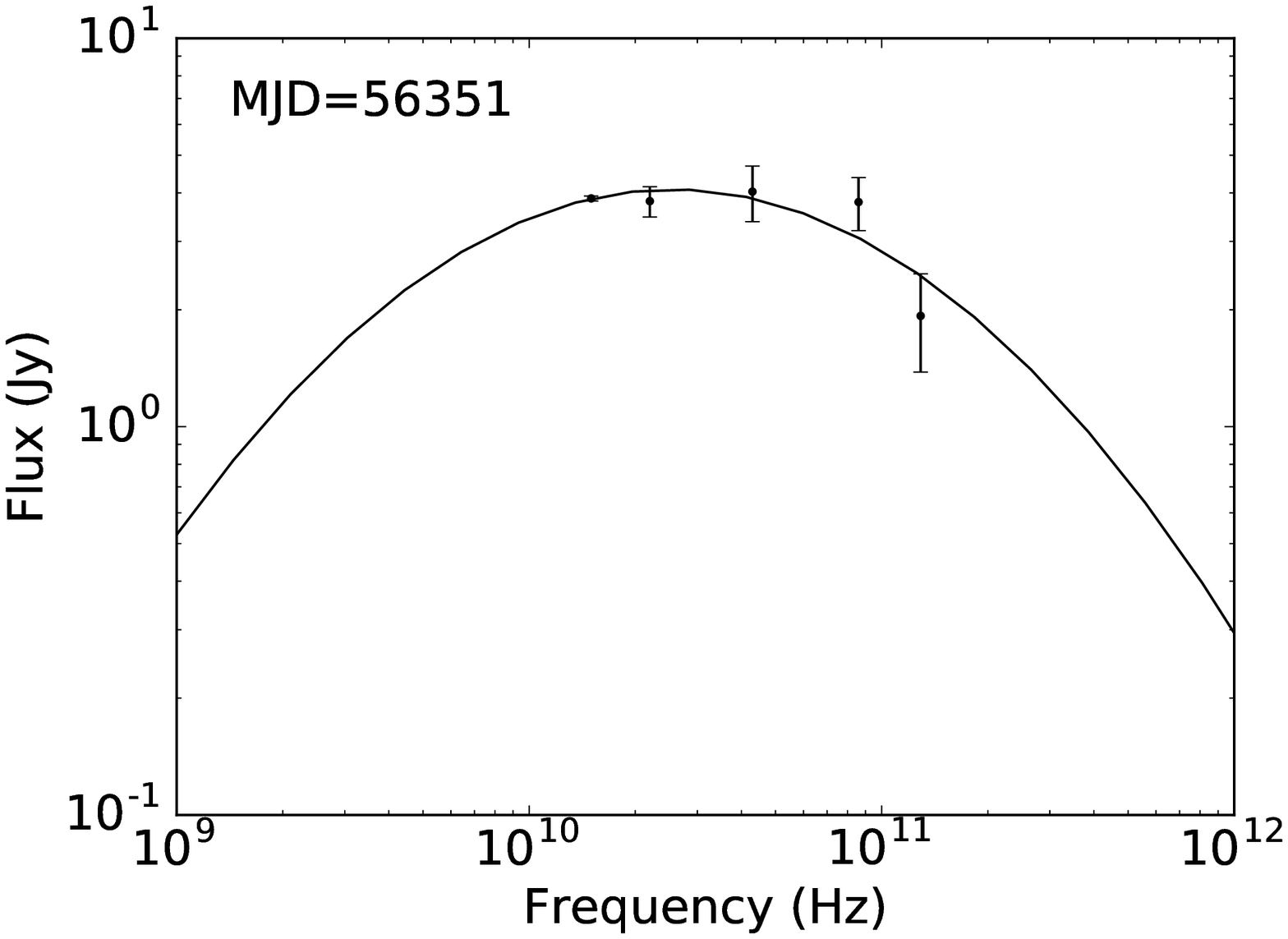}
\includegraphics[height=4cm, width=6cm,trim=0cm 0cm 0cm 0cm,clip=true]{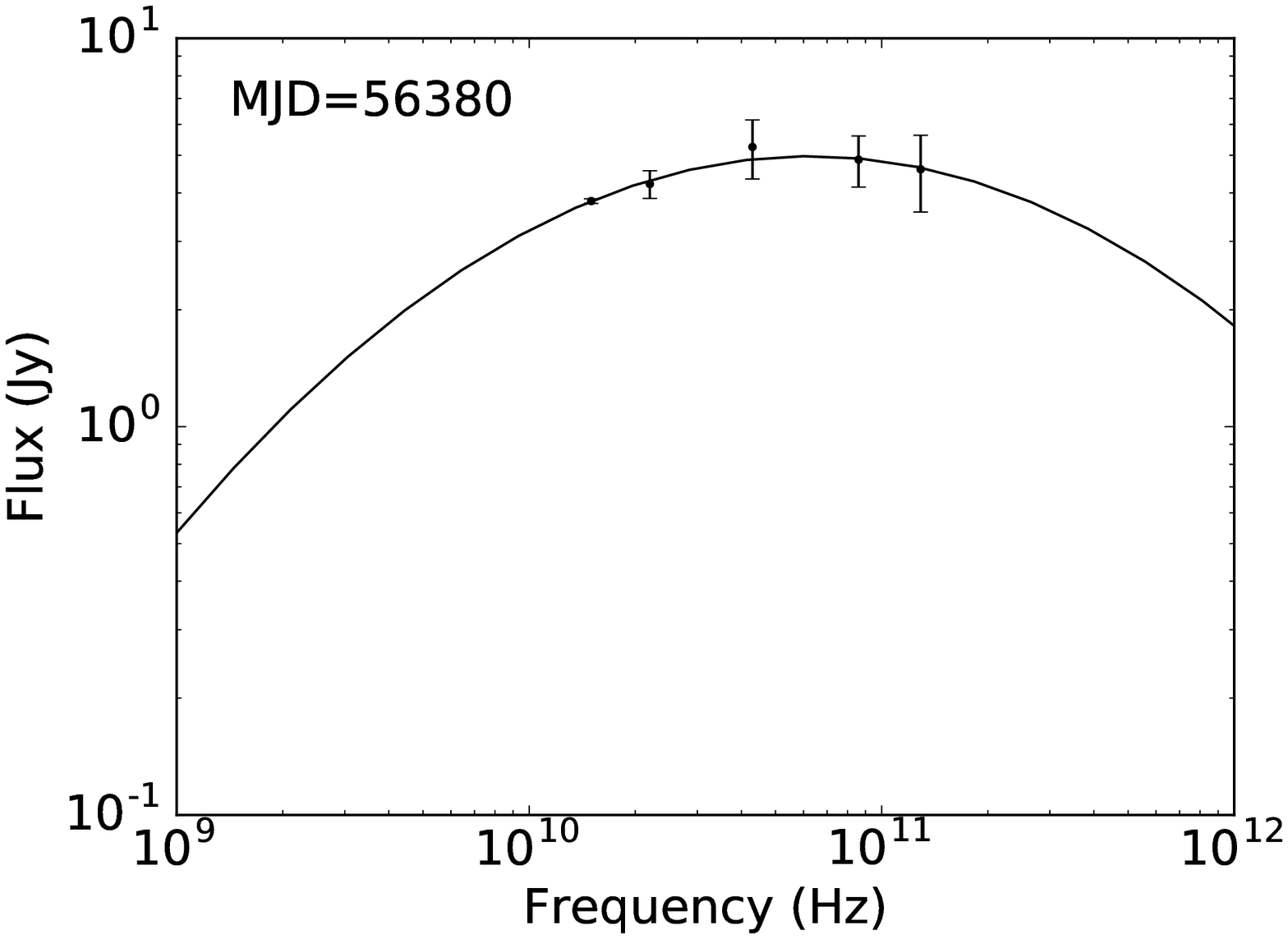}\\
\includegraphics[height=4cm, width=6cm,trim=0cm 0cm 0cm 0cm,clip=true]{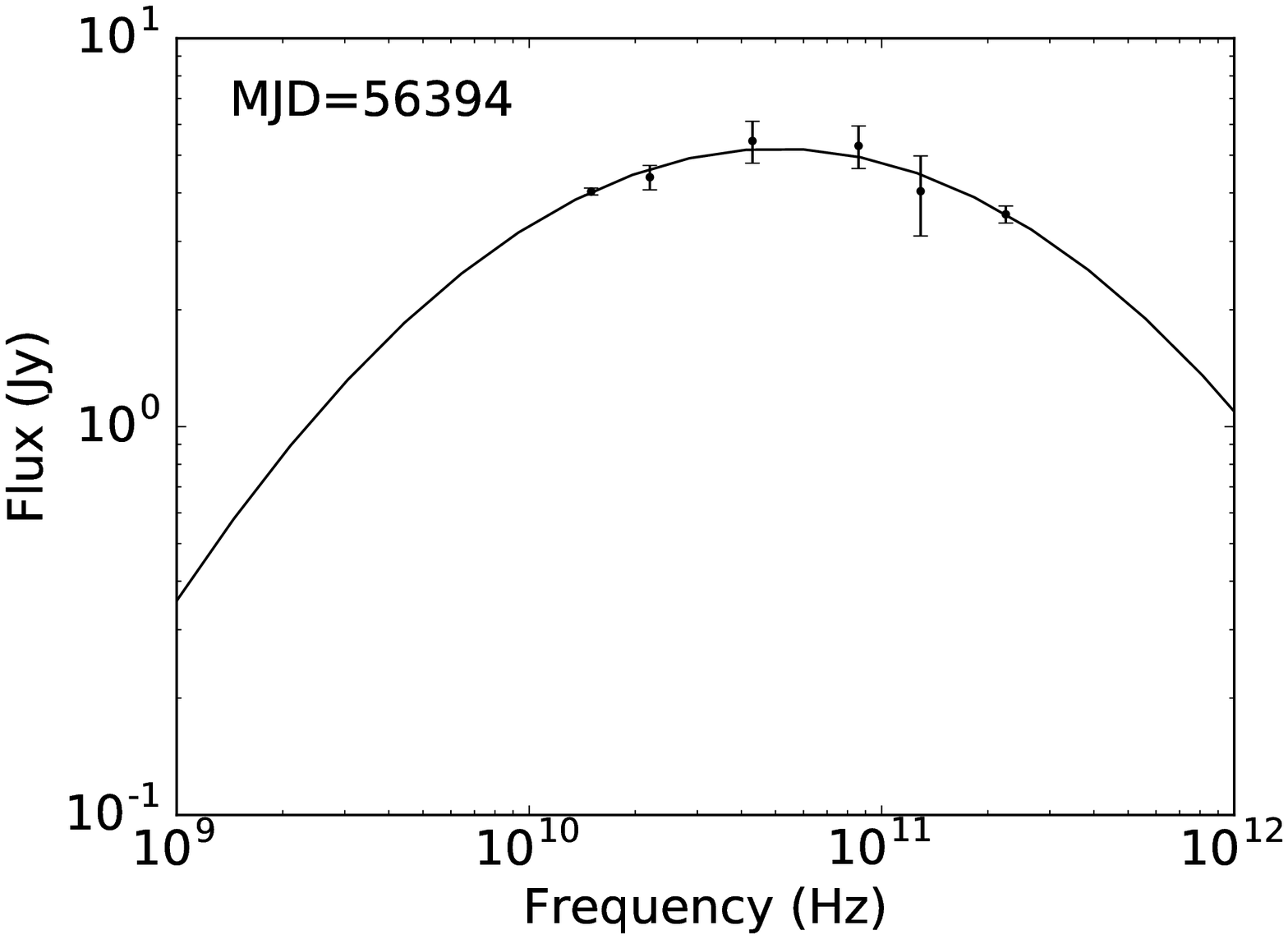}
\includegraphics[height=4cm, width=6cm,trim=0cm 0cm 0cm 0cm,clip=true]{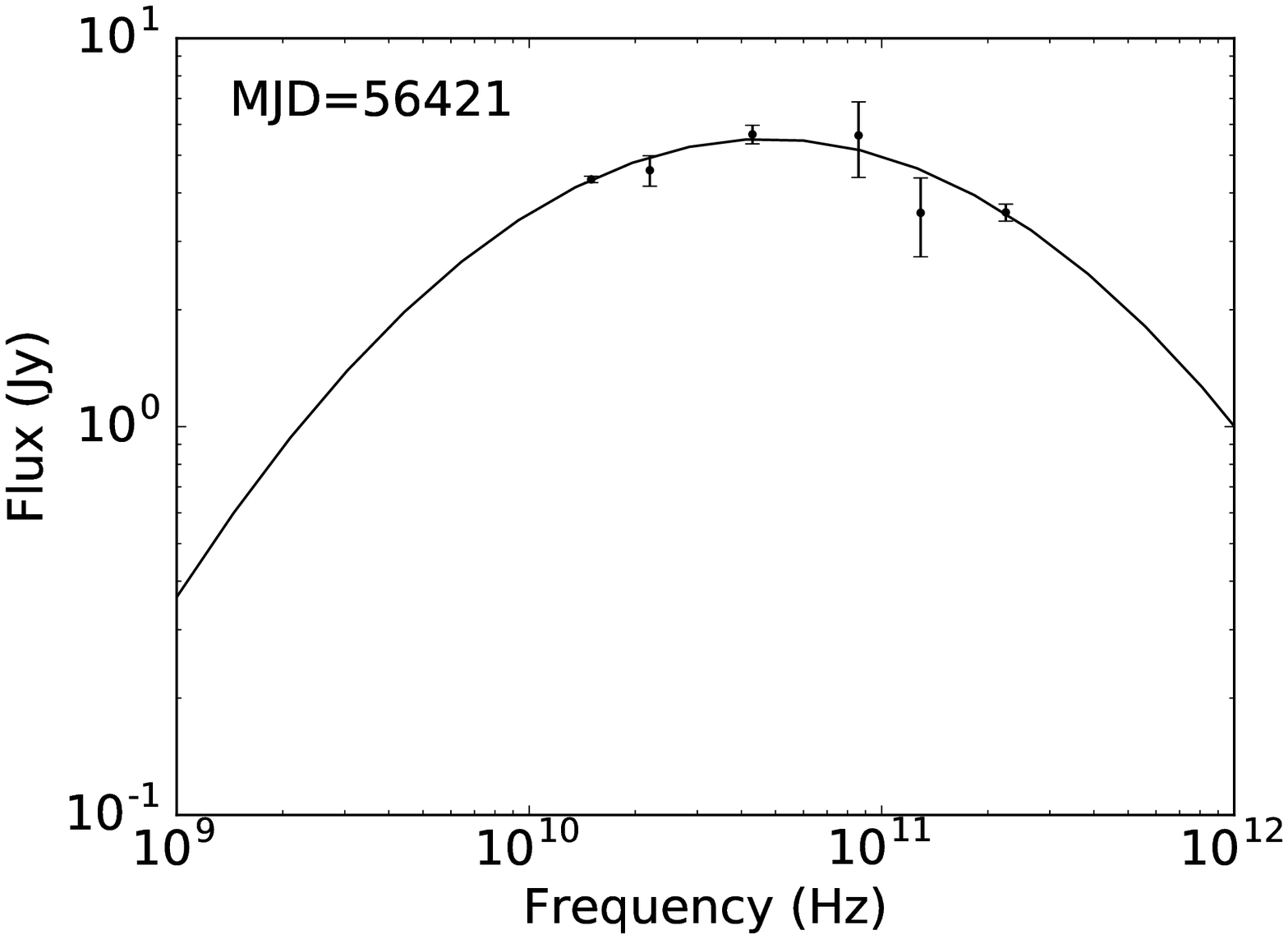}
\includegraphics[height=4cm, width=6cm,trim=0cm 0cm 0cm 0cm,clip=true]{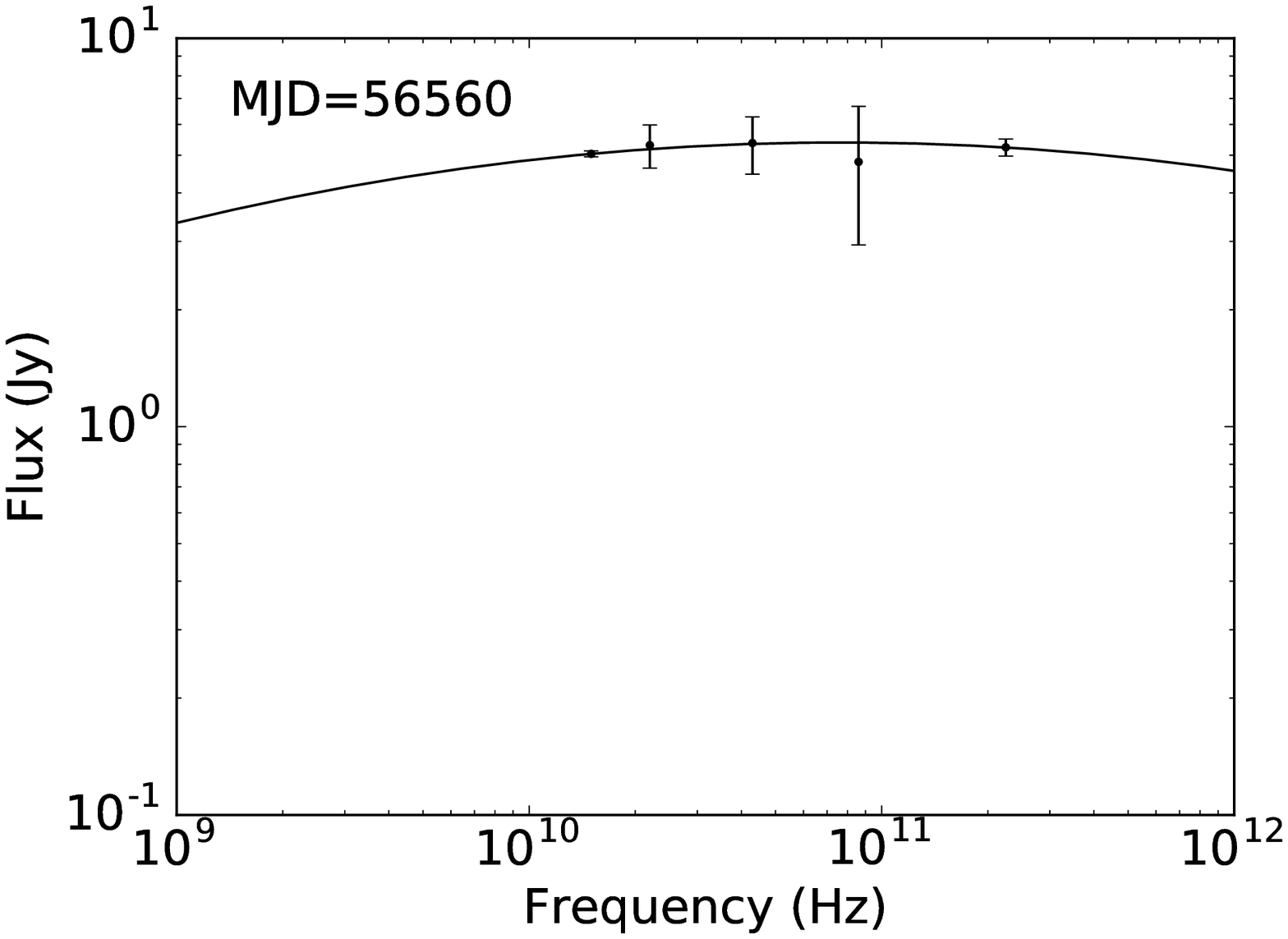}\\
\includegraphics[height=4cm, width=6cm,trim=0cm 0cm 0cm 0cm,clip=true]{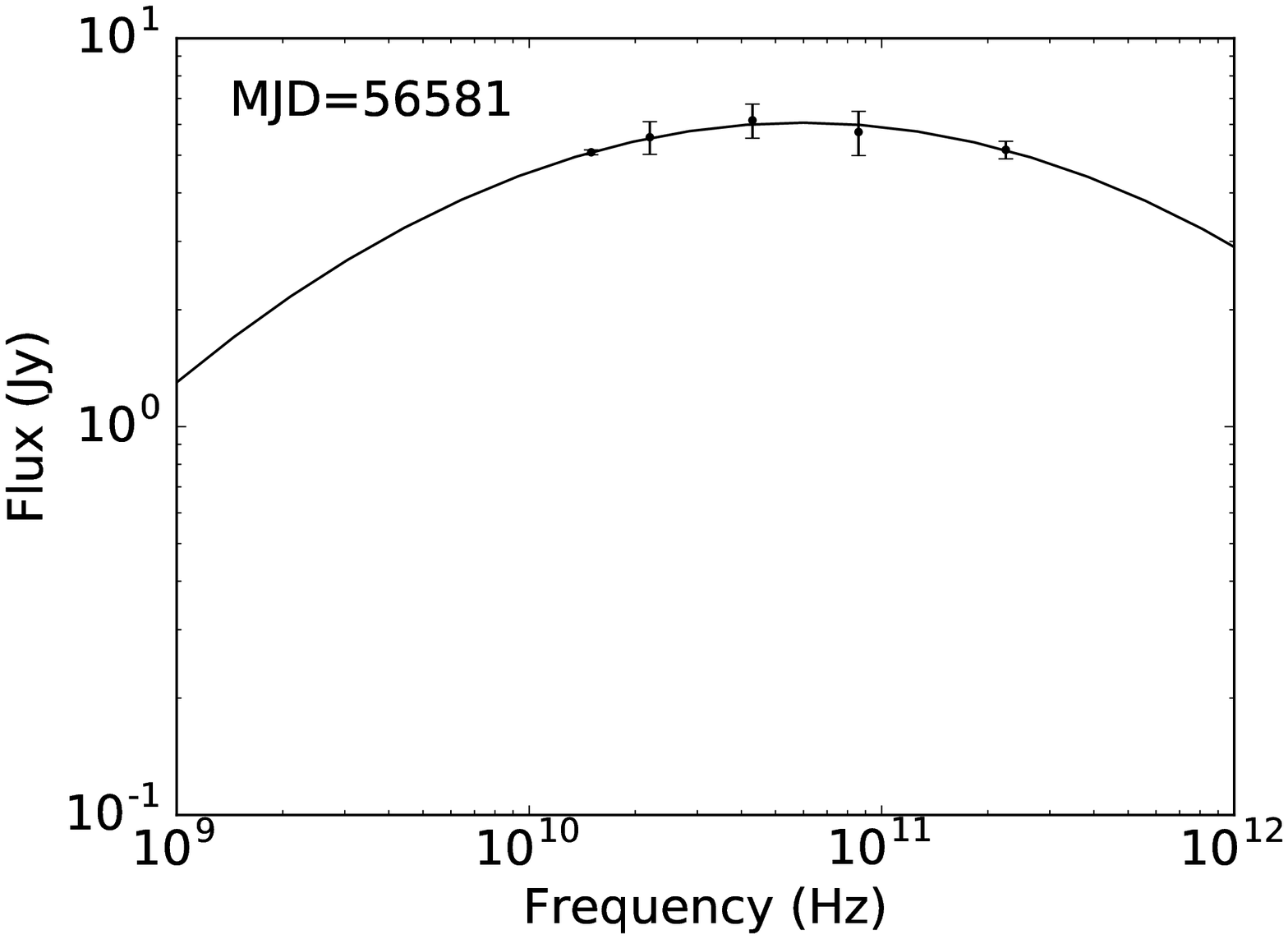}
\includegraphics[height=4cm, width=6cm,trim=0cm 0cm 0cm 0cm,clip=true]{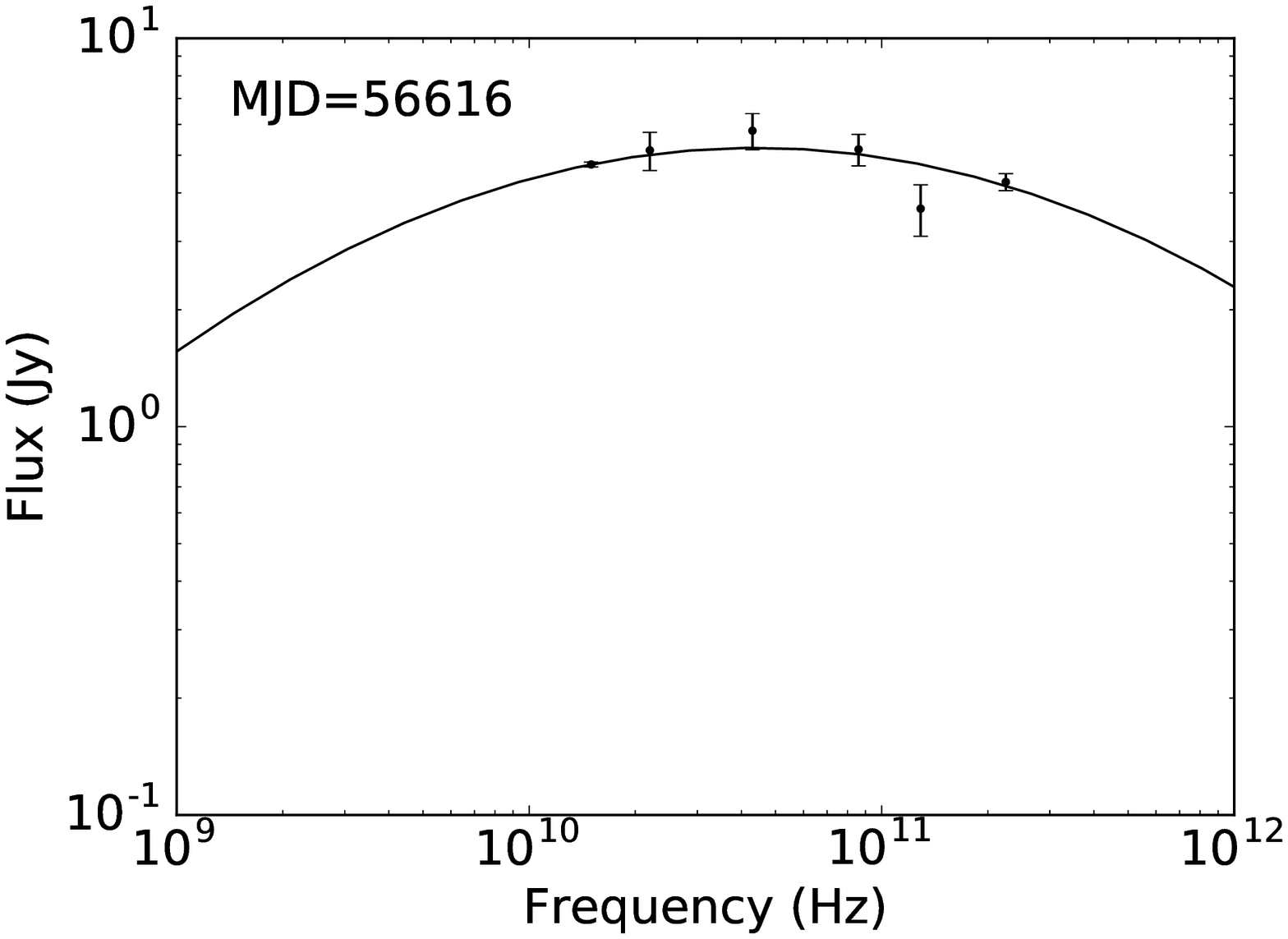}
\includegraphics[height=4cm, width=6cm,trim=0cm 0cm 0cm 0cm,clip=true]{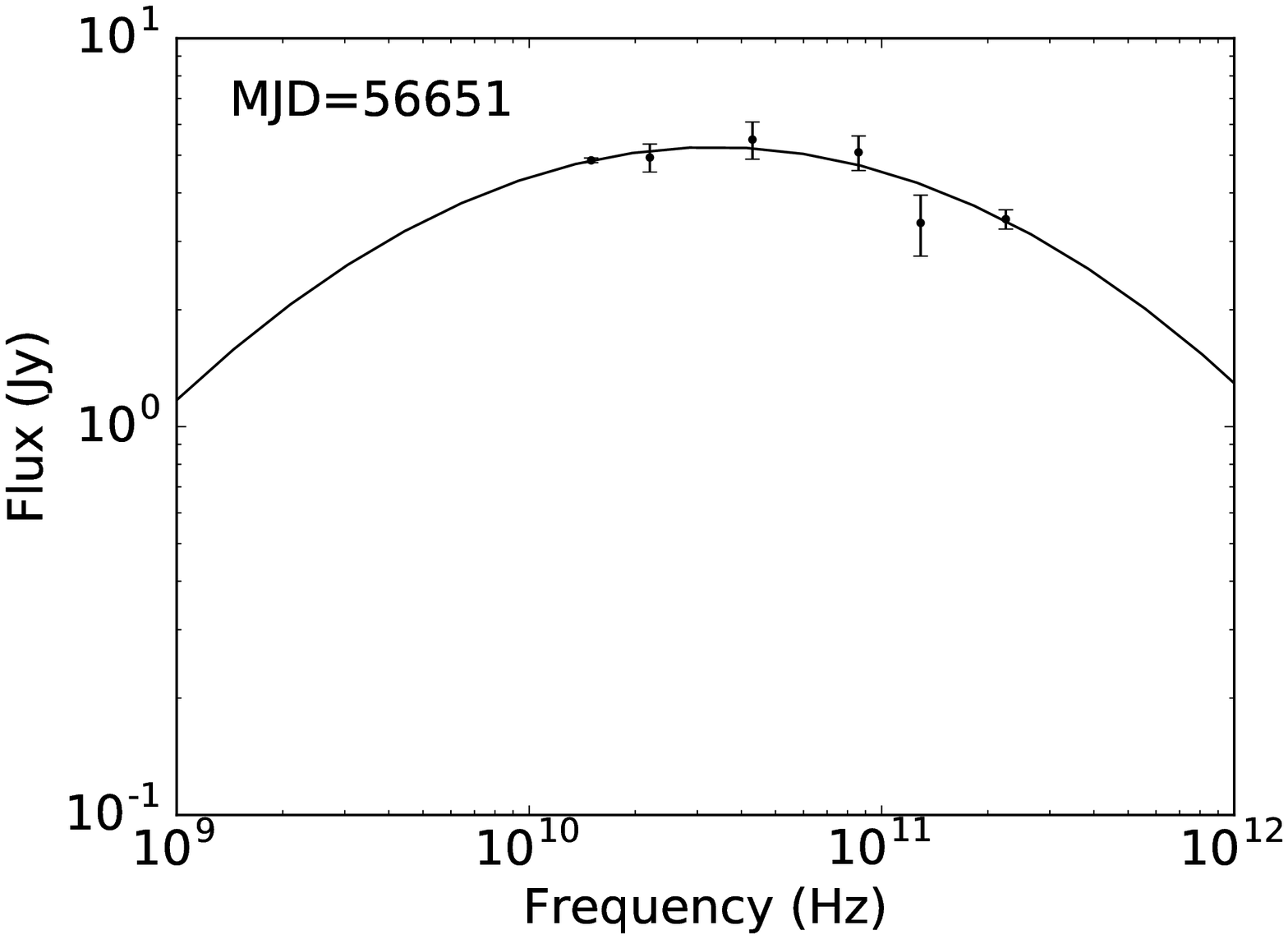}\\
\includegraphics[height=4cm, width=6cm,trim=0cm 0cm 0cm 0cm,clip=true]{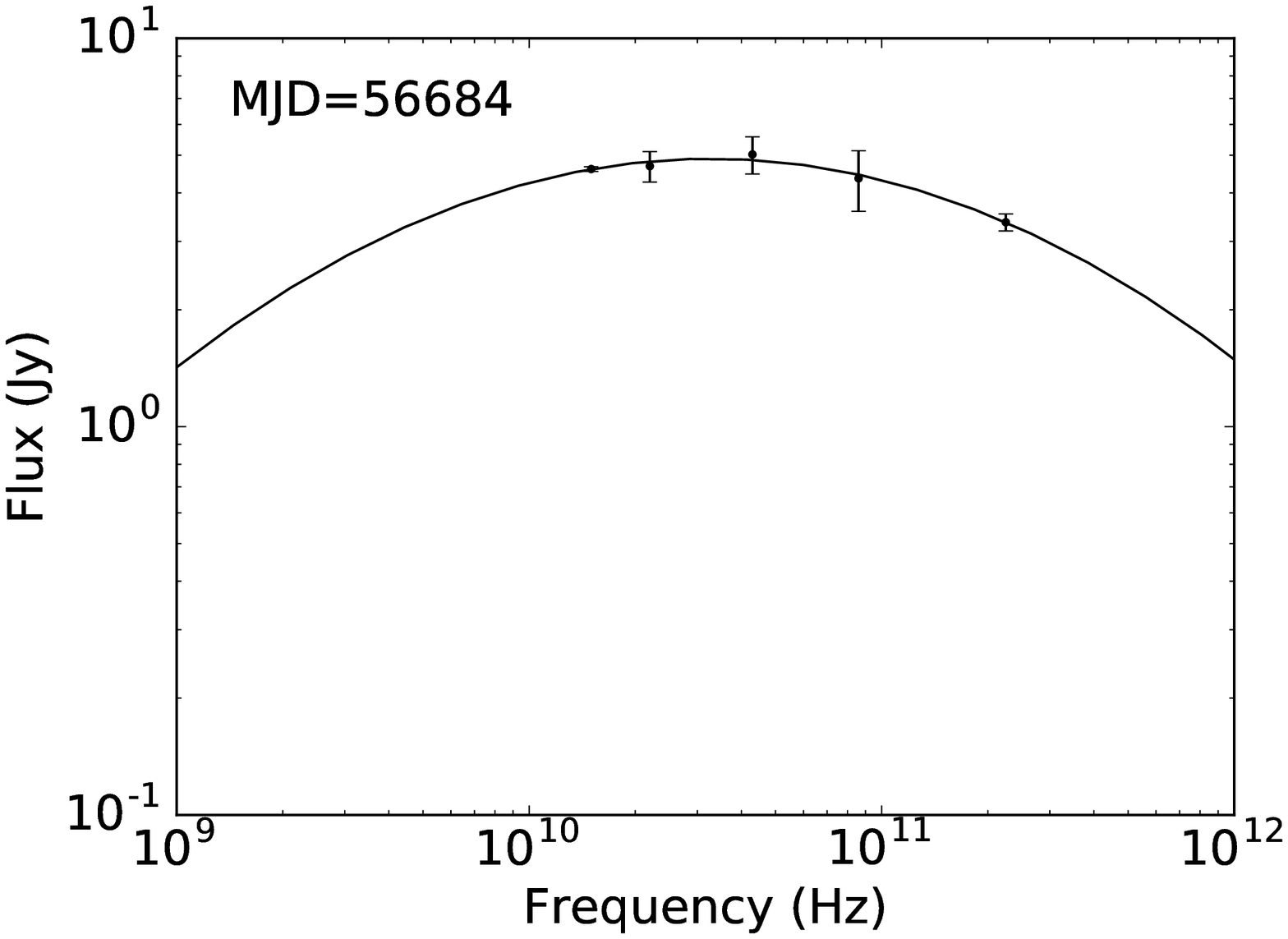}
\includegraphics[height=4cm, width=6cm,trim=0cm 0cm 0cm 0cm,clip=true]{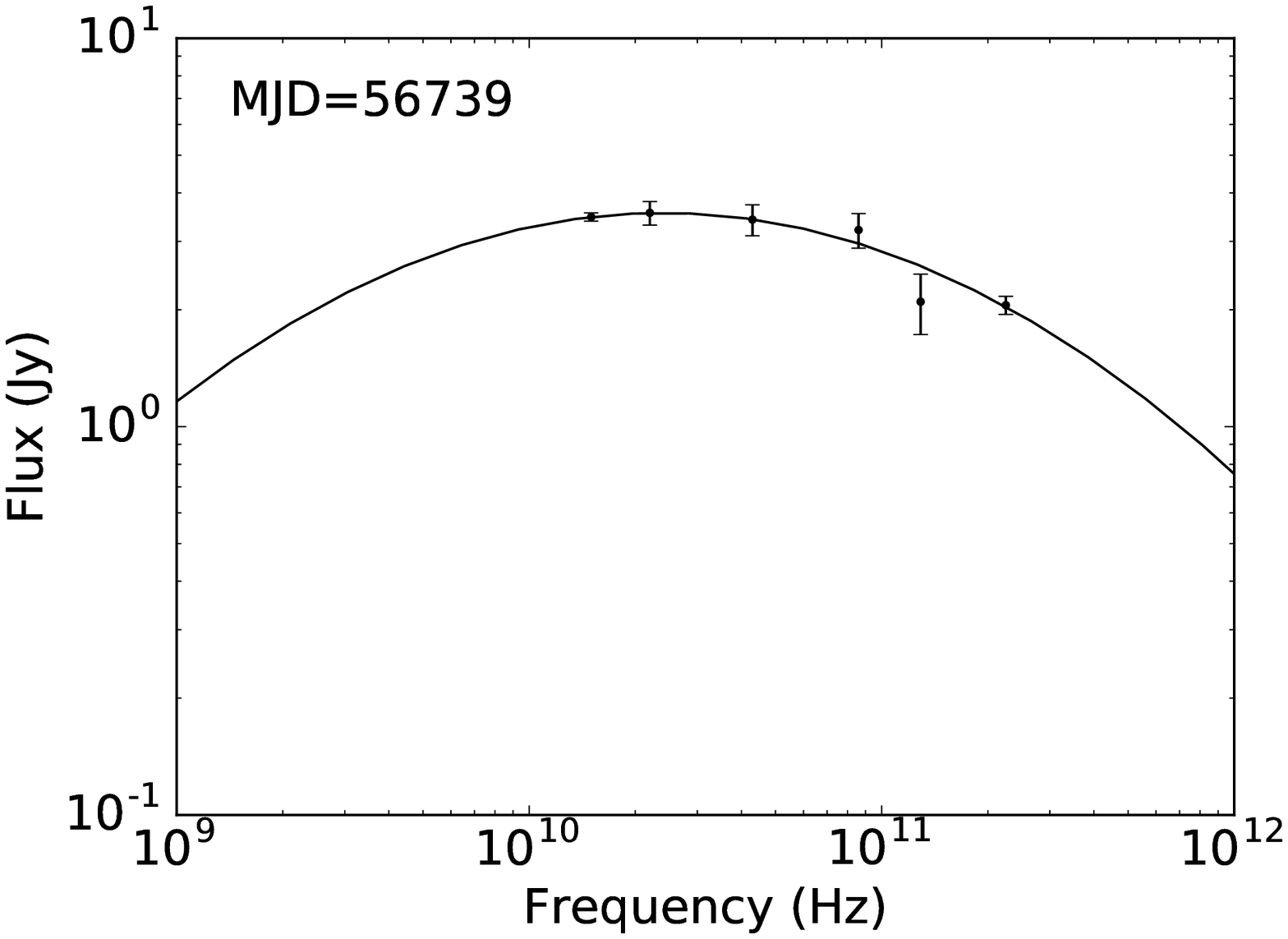}
\includegraphics[height=4cm, width=6cm,trim=0cm 0cm 0cm 0cm,clip=true]{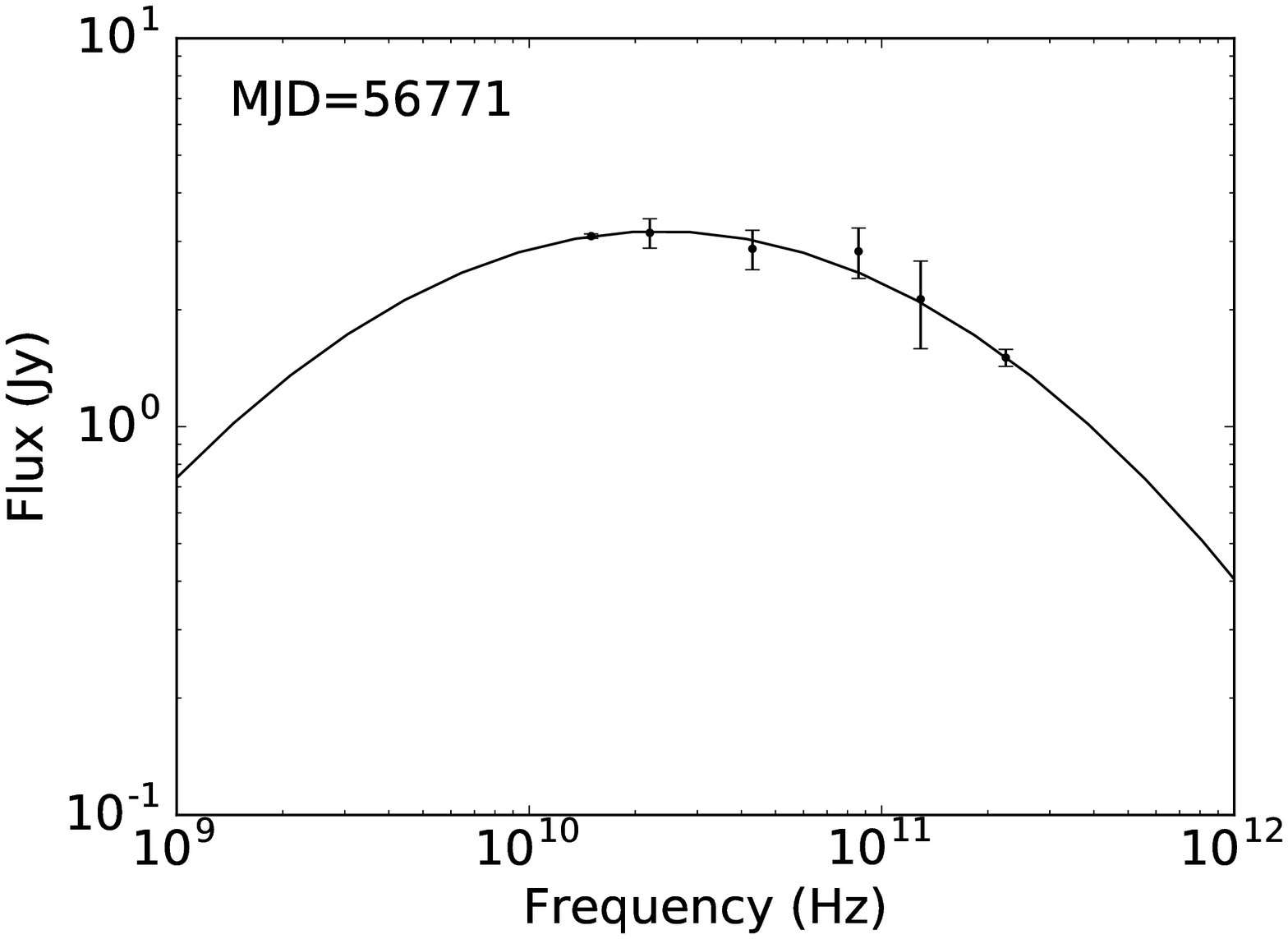}\\
\includegraphics[height=4cm, width=6cm,trim=0cm 0cm 0cm 0cm,clip=true]{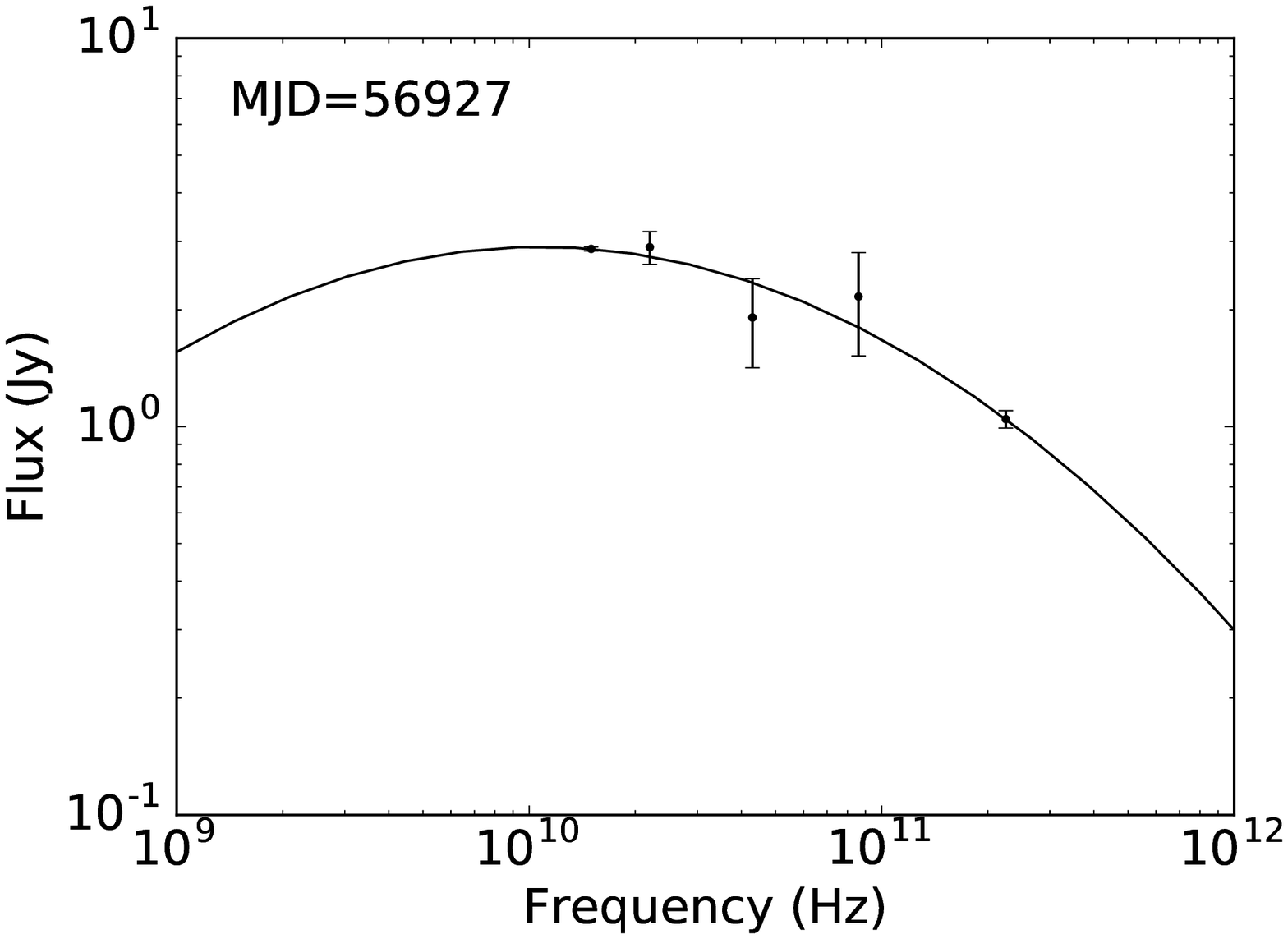}
\includegraphics[height=4cm, width=6cm,trim=0cm 0cm 0cm 0cm,clip=true]{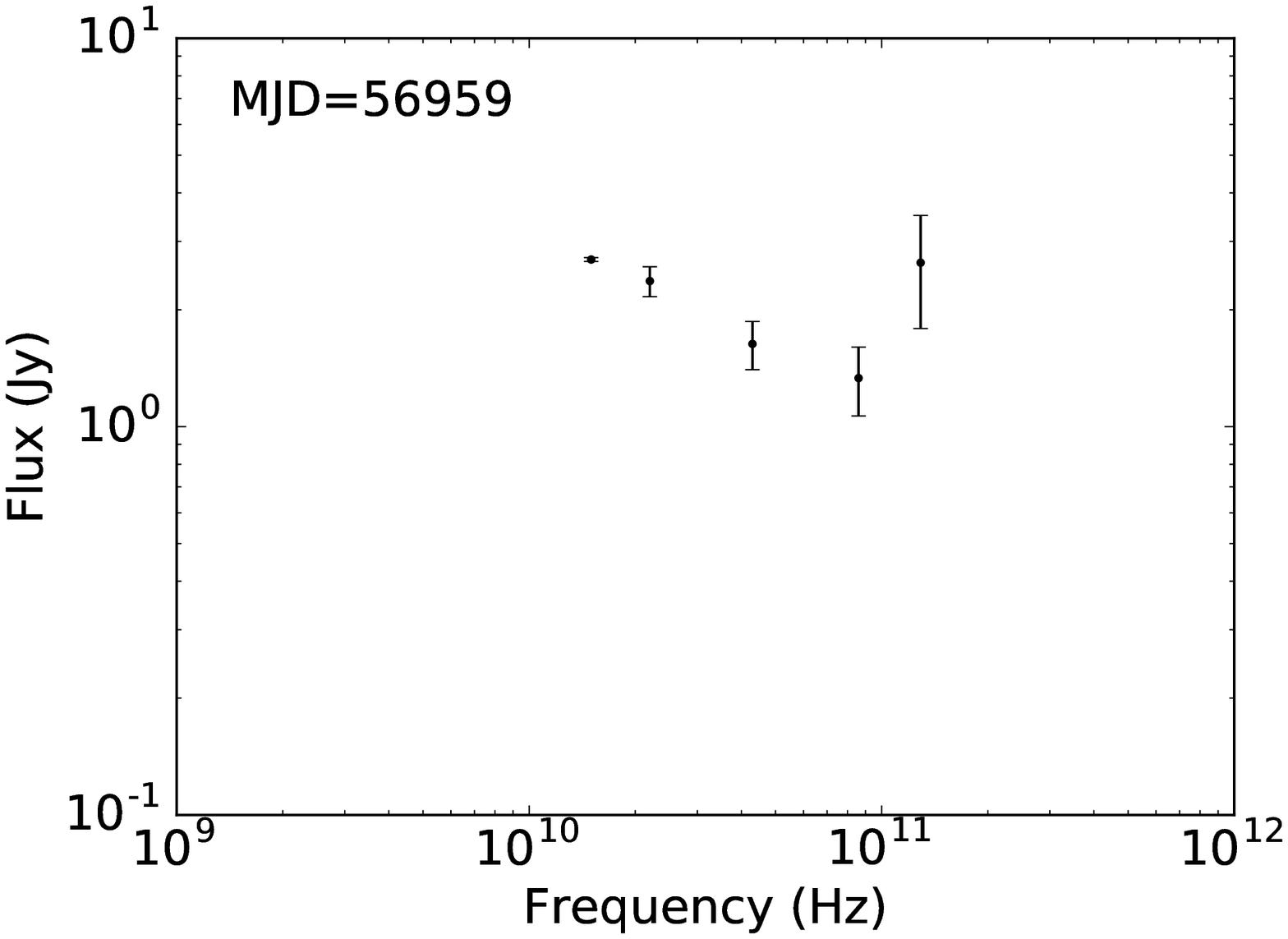}
\includegraphics[height=4cm, width=6cm,trim=0cm 0cm 0cm 0cm,clip=true]{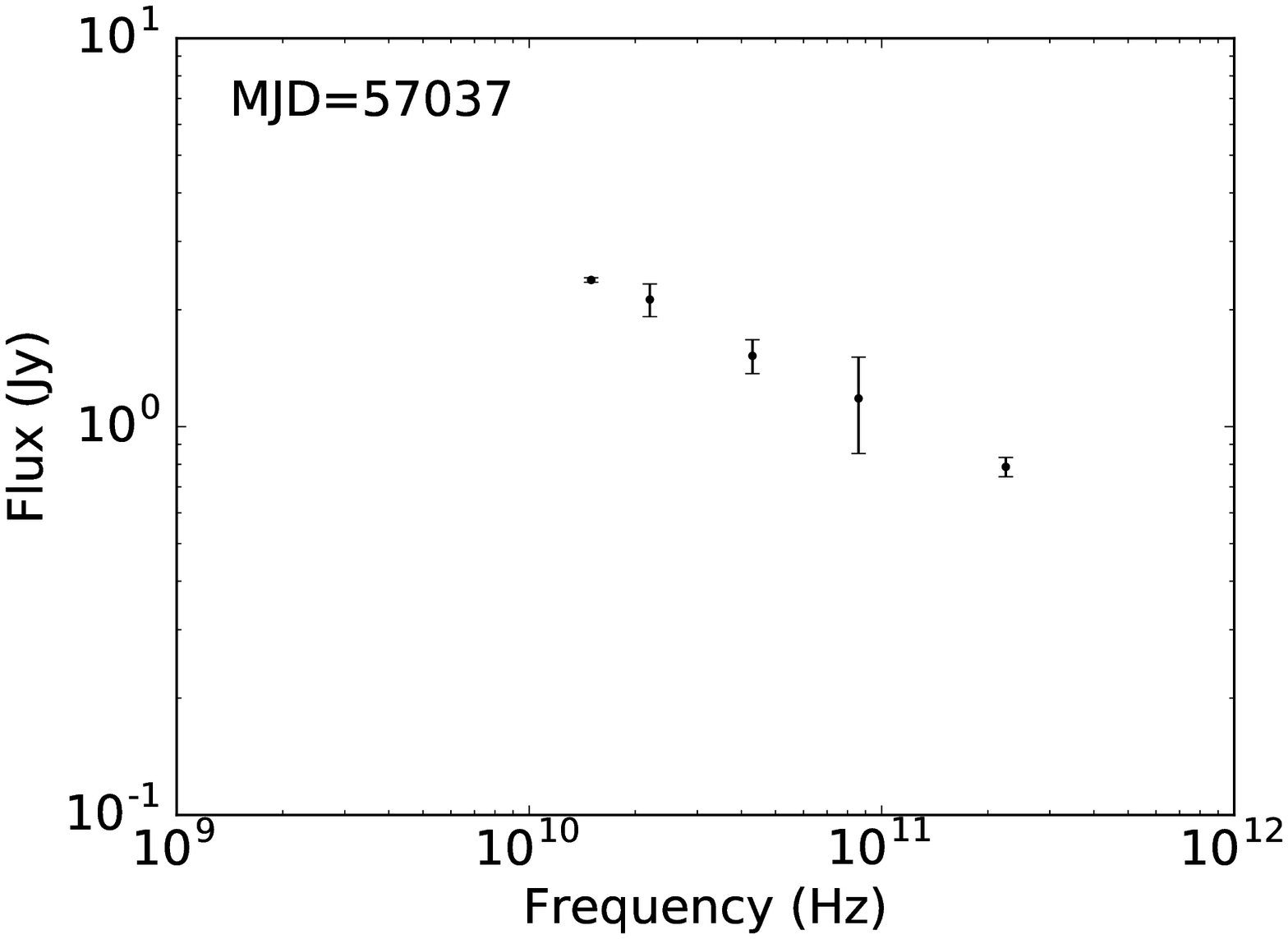}\\
\includegraphics[height=4cm, width=6cm,trim=0cm 0cm 0cm 0cm,clip=true]{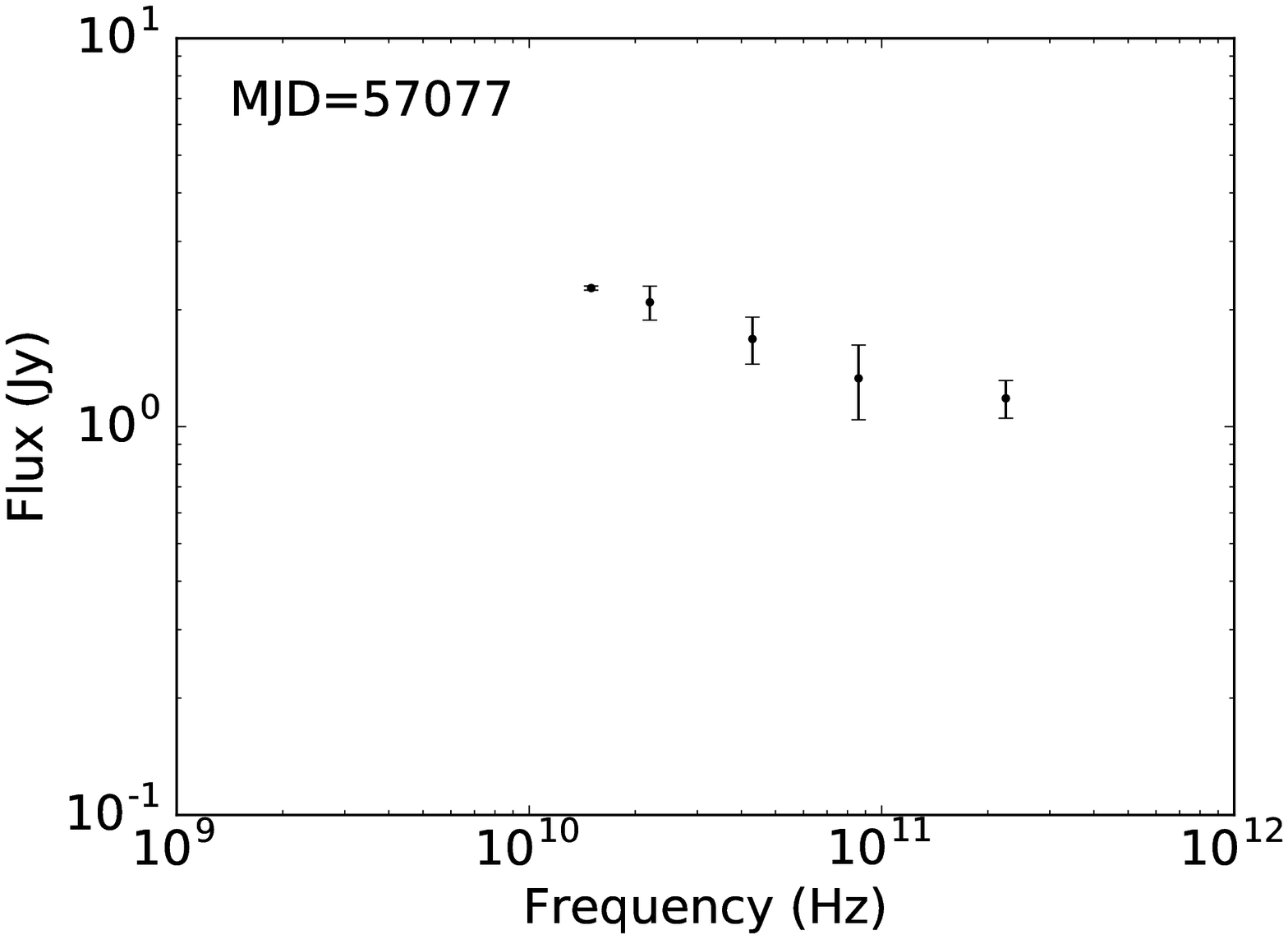}
\includegraphics[height=4cm, width=6cm,trim=0cm 0cm 0cm 0cm,clip=true]{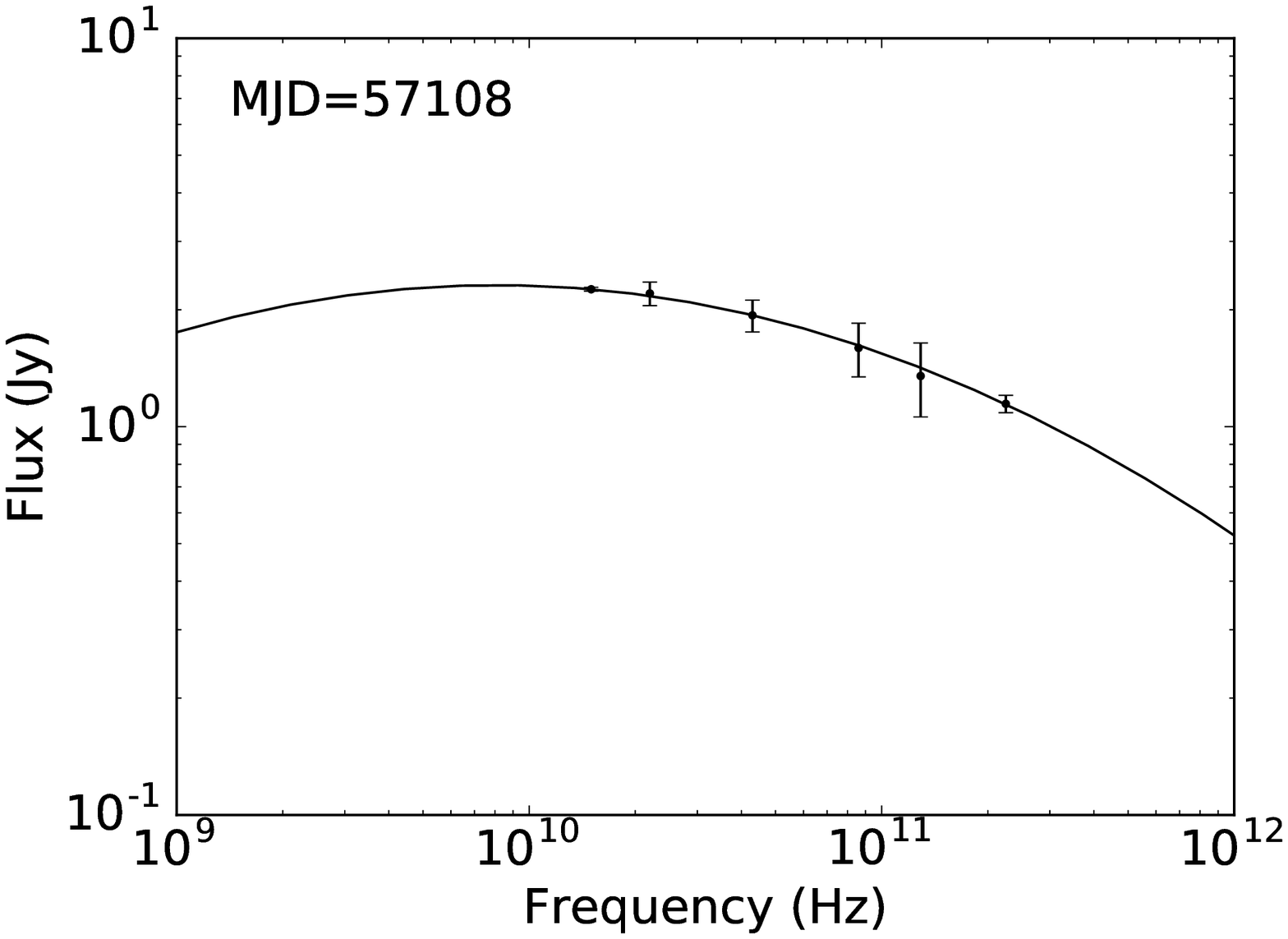}
\caption{Radio SED for the various epochs for which quasi--simultaneous (within 14 days) radio data from 15 to 225~GHz was available. Epoch is indicated in the top left corner. Whenever a fit of the form given in the text was possible, it is shown by the curve.}
\label{SED}
\end{figure*}

In Figure \ref{SED} we show the above described SEDs. The first noticeable characteristic is that, despite 1633+382 being a relatively flat-spectrum AGN, a certain curvature can be seen in the SED for certain epochs. As discussed in Paper~I, as a general trend, the source seems to be slightly optically thick at lower radio frequencies, whereas it progressively becomes optically thin at higher radio frequencies, leading to an observed turnover frequency $\nu_c$ that can be measured. This behaviour is not constant with time and the spectral indices vary at different epochs, following a certain correlation with the flux density enhancements (see Figure 6 of Paper~I). As can be seen in Figure \ref{SED}, this leads to a change in  $\nu_c$. 

In order to study this in detail, we fitted the SED with a function of the form $S=S_0\left(\nu/\nu_c\right)^{-a\log(\nu/\nu_c)}$\citep{Sambruna96,Massaro04,Rani11,Prince17}, where $a$ can be considered to be the spectral index at large frequencies after the turnover. Such log-parabolic distribution is not only a simple mathematical tool for spectral modeling. As \cite{Massaro06} suggest, under the assumption of electrons accelerated via a two-step process with a broken power-law energy distribution, the resulting electron spectrum can be described by a log-parabola when cooling effects are considered. We favor this function over some other functions such a broken power law due to its simplicity and better ability of convergence to our data. We were able to fit 13 epochs in this way, although in four epochs the fit failed to converge due to the difficulty of finding any turnover frequency with the current data. As a check for consistency, we repeated this procedure excluding the OVRO 15~GHz data and using only iMOGABA data, finding in both cases consistent results in the fitted parameters, with the possible exception of epoch 56560, for which the significant flatness of the spectrum produced a large uncertainty in the fitted values. In Figure \ref{SED} we show the spectral fits to the data.

\begin{figure}
\includegraphics[scale=0.43,trim={0cm 0cm 0cm 0cm},clip]{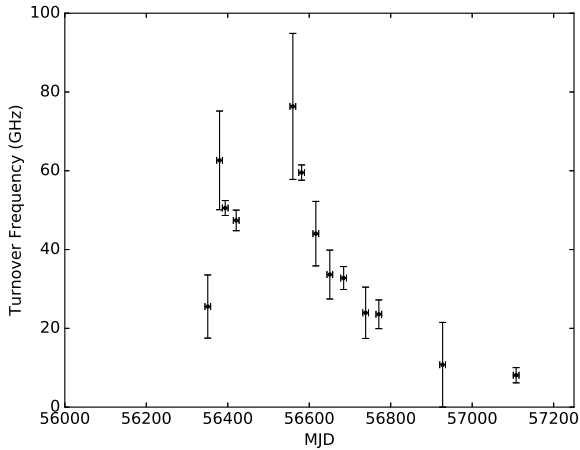} 
\caption{Radio turnover frequencies as defined in fitting equation for all the epochs where the fit was possible. }
\label{Turnover}
\end{figure}

Turnover frequencies estimated from the spectral fits are shown in Table \ref{turnovertable}. In Figure \ref{Turnover}, we plot these as a function of their corresponding epochs. As anticipated from the qualitative discussion above, it is clear that the obtained values are inconsistent with a constant turnover frequency over time. Instead, a change by at least a factor of 3 in the value of $\nu_c$ can be seen. 

\begin{figure}
\includegraphics[scale=0.43,trim={0cm 0cm 0cm 0cm},clip]{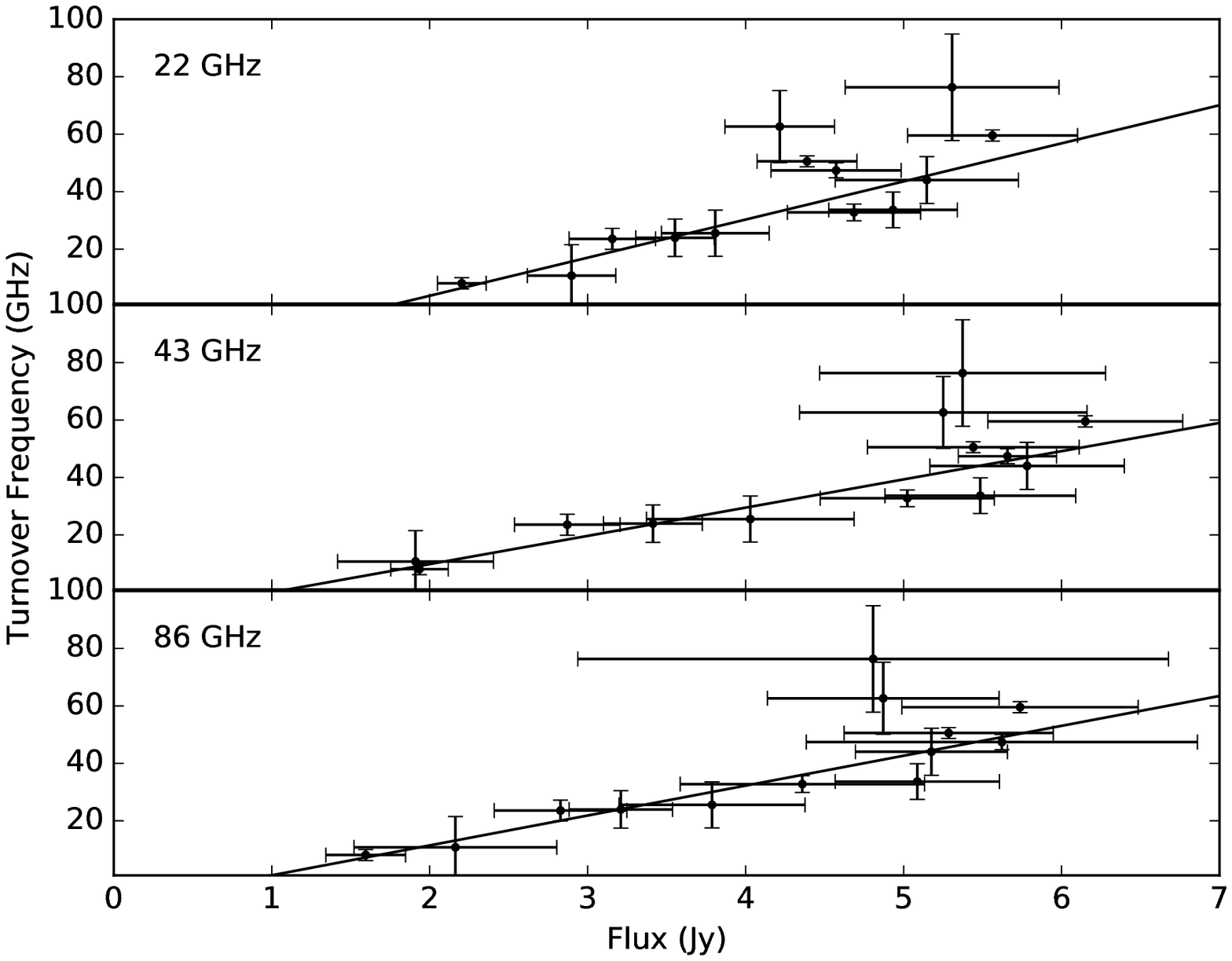}
\caption{Radio turnover frequencies a function of the iMOGABA flux densities at 22, 43 and 86~GHz. Thin black line indicates the best linear fit.}
\label{Turnoverflux}
\end{figure}

It is quite remarkable that $\nu_c$ seems to follow a similar trend than that of the light curves, i.e., the larger the flux density, the higher the turnover frequency.  In order to examine this in more detail, in Figure \ref{Turnoverflux} we plot $\nu_c$ as a function of the iMOGABA flux densities at 22, 43 and 86~GHz. It appears that, for low flux densities, there is a very well defined linear relation that seems to be broken at high $\nu_c$, where it seems that a saturation flux density level is reached and an increase of $\nu_c$ is not followed by an equivalent increase of flux density anymore. This seems to be very well related with characteristic shock model SED evolution, as we will extensively describe below.

\begin{table}
\begin{center}
\caption{Turnover frequency and Magnetic Fields}
\label{turnovertable}
\begin{tabular}{cccc}
\tableline
\hline
Epoch (MJD) & $\nu_c$ (GHz) & $B_{SSA}$ (mG)&$B_{eq}$ (mG)\\
(1)&(2)&(3)&(4)\\
\tableline
56351	&	$25\pm8$		&$	0.03	\pm	0.02	$&$	357	\pm	35	$\\
56380	&	$62\pm12$	&$	0.19	\pm	0.13	$&$	706	\pm	68	$\\
56394	&	$51\pm2$		&$	0.12	\pm	0.08	$&$	599	\pm	57	$\\
56421	&	$47\pm3$		&$	0.12	\pm	0.08	$&$	547	\pm	52	$\\
56560	&	$76\pm18$	&$	0.06	\pm	0.04	$&$	1146	\pm	109	$\\
56581	&	$59\pm2$		&$	0.02	\pm	0.01	$&$	1035	\pm	99	$\\
56616	&	$44\pm8$		&$	0.01	\pm	0.01	$&$	854	\pm	82	$\\
56651	&	$34\pm6$		&$	0.01	\pm	0.01	$&$	620	\pm	59	$\\
56684	&	$33\pm3$		&$	0.02	\pm	0.02	$&$	510	\pm	49	$\\
56739	&	$24\pm7$		&$	0.06	\pm	0.04	$&$	294	\pm	29	$\\
56771	&	$24\pm4$		&$	0.06	\pm	0.04	$&$	296	\pm	29	$\\
56927	&	$11\pm11$	&$	0.01	\pm	0.01	$&$	158	\pm	16	$\\
57108	&	$8\pm2$		&$	0.01	\pm	0.01	$&$	117	\pm	11	$\\

\tableline
\end{tabular}
\end{center}
\end{table}

\subsection{Evolution of Radio Fluxes}

If we adopt the shock--in--jet model \citep{MarscherGear85}, we can consider the spectral evolution of a flare in the turnover frequency -- turnover flux density ($S_{\nu_c} - \nu_c$) plane. The typical evolution can be divided into three different stages \citep[see e.g.][for more details]{Fromm11}. During the first stage, the Compton losses are dominant and $S_{\nu_c}$ decreases, while $\nu_c$ increases. In the second stage, synchrotron losses become dominant and $S_{\nu_c}$ becomes roughly constant while $\nu_c$ decreases. In the adiabatic stage, both $S_{\nu_c}$ and $\nu_c$ decrease. Thus, in the standard model, we can parametrize $S_{\nu_c}\propto \nu_c^{\epsilon_i}$, with $\epsilon_i$ for each of these stages depending on physical quantities such as the magnetic field, Doppler factor or energy of the relativistic electrons. 

The synchrotron stage has a flat evolution of $S_c$ vs. $\nu_c$ only in the case considered by \cite{MarscherGear85} of a viewing angle small enough that our line of sight crosses the narrow part of the shock (in the aberrated frame). The other case, with a wider viewing angle, was considered by \cite{Bjornsson00}, and $S_c$ is proportional to (approximately) $\nu_c^{0.8}$, so $\nu_c$ increases as the flare rises. For synchrotron self-Compton (SSC), \cite{MarscherGear85} suggested $\epsilon_{Compton}=-2.5$, while \cite{Fromm11} suggested $\epsilon_{Compton}=-1.2$, assuming that the flaring flux density is produced by accelerated particles within a small layer behind the shock front, with its width depending on the dominant cooling process, and \cite{Bjornsson00} favored $\epsilon_{Compton}=-0.4$. For external Compton (EC), which is generally favored for gamma-ray bright quasars, the evolution will be softer, so that the spectrum will not rise so fast and $\epsilon_{EC}$ will be closer to zero. One would however need to know how the external seed photon density decreases with distance down the jet in order to predict $\epsilon_{Compton}$. Finally, predicted values for the adiabatic stage range from $\epsilon_{adiabatic}=0.69$ \citep{MarscherGear85} to $\epsilon_{adiabatic}=0.77$ \citep{Fromm11}. We note that the inverse relationship between $S_{\nu_c}$ and $\nu_c$ is quite generic for an expanding (unbeamed) synchrotron source, as discussed in \cite{vanderLaan66}.

\begin{figure}
\includegraphics[scale=0.43,trim={0cm 0cm 0cm 0cm},clip]{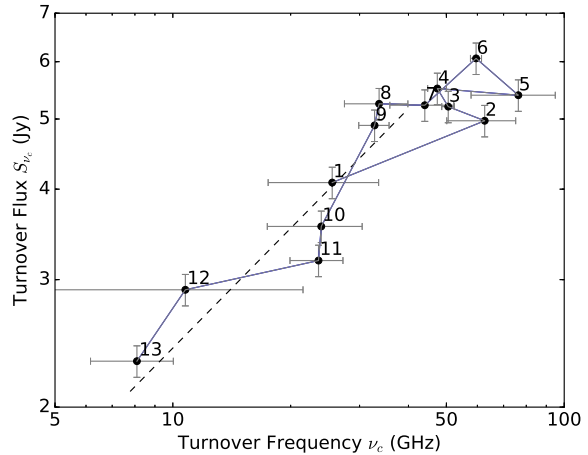}
\caption{Time evolution of $S_{\nu_c}$ vs.\  $\nu_c$ for the radio flux enhancement. Numbers indicate the various correlative different epochs for which $\nu_c$ was calculated (see Fig.\ \ref{Turnover}). Blue line connects the different epochs for eye guidance. Dashed black line indicates the fit $\epsilon=0.6\pm0.1$ for points 8--13 (see text). Note that the time gap between two consecutive numbers (=epochs) may be different.} 
\label{fig-hysteresis}
\end{figure}

We plot the evolution in the ($S_{\nu_c} - \nu_c$) plane for the flux density enhancements shown by 1633+382 for the observed period in Figure \ref{fig-hysteresis}.  We indicate the various epochs listed in Table \ref{turnovertable} for which we obtained turnover frequency values by consecutive numbers (i.e., 1= MJD56351; 2=MJD56380; etc) for eye guidance. We note that the time gap between two consecutive numbers (=epochs) may be different. The first data point corresponds to the epoch for which the $\gamma-$ray flux is around a local minimum between two flares. As a reminder, the radio flux density is estimated to be delayed by about $\sim90$~days (see Paper~I). The trajectory of the flux enhancement in the ($S_{\nu_c} - \nu_c$) plane is subject to uncertainties and limited cadence, and thus the discussion will not be robust, but we can still suggest a possible interpretation.

Points 1--4 seem to be related with the first radio flux enhancement. There seems to be a period (points 2--4) where $\nu_c$ decreases while $S_c$ rises slightly. This looks  like the synchrotron stage (for which $\epsilon\sim0$), or it could be an EC Compton stage. Points 5 onwards seem to be related with the second flux enhancement. Points 5--8 indicate a stage where $S_c$ stays nearly constant while $\nu_c$ decreases, which is consistent with synchrotron decay for a small viewing angle. This suggests a viewing angle $\theta<1/\Gamma$, in agreement with the considerations $\theta\sim2.5\degr$ and $\Gamma\sim12-14$ above. In points 8--13, both $S_{\nu_c}$ and $\nu_c$ decrease, suggesting the adiabatic decay stage. The estimated $\epsilon_{adiabatic}=0.6\pm0.1$ is compatible with the predictions. This suggests that a simple model with a constant Doppler factor may be applicable for this source. This may also be indicated by the apparent constant speed of the two ejected components C2 and C3 (see Figure  \ref{fig_components}b). By comparison, values found in 0716+714 \citep{Rani13b} were much larger than the theoretical estimations, suggesting that a simple model with constant Doppler factor may not be the case in that source.

\subsection{Magnetic Fields}

If we consider that the turnover frequency variations are due to synchrotron self-absorption, we can estimate the magnetic field in a homogeneous, incoherent synchrotron radio source with a power--law electron energy distribution with \citep[e.g.][]{Jones74,Kellerman81,Marscher83} 
\begin{equation}
B_{SSA}\sim3.8\times10^{-5} \Theta_c^4 \nu_c^5 S_{\nu_c}^{-2}\left(\frac{\delta}{1+z}\right)^{-1}, 
\label{eq_B_SSA}
\end{equation}
where  $\Theta_c$ is the size of the component in mas, $\delta$ is the Doppler factor and $S_{\nu_c}$ is the flux density in Jansky at the turnover frequency $\nu_c$ in GHz. 
This expression is different from that in \cite{Marscher83}, which applies to moving features, since we consider the VLBI core to be in a roughly steady state rather than evolving in time. As a consequence, one of the $\delta$ factors from the flux density transformation into the observer's frame is removed. This changes the derivation of the magnetic field when one observes the SSA turnover in the spectrum. In this case, the $\delta/(1+z$) factor is raised to the $-1$ power rather than $+1$ power.  
To estimate $\Theta_c$, we interpolated the sizes obtained from the BU-VLBI program to the epochs where $S_{\nu_c}$ was obtained. Then we calculated the interpolated size from 43~GHz to the $S_{\nu_c}$ value assuming $\Theta_c\propto\nu^{-0.7}$, which represents a semi-parabolic geometry (we will discuss this geometry in a forthcoming paper). We checked that the results do not change significantly if we consider the pure conical or pure parabolic cases. Even though there is a number of assumptions and approximations involved, we consider the value obtained in this way is more reliable and adequate.

\begin{figure}
\includegraphics[scale=0.43,trim={0cm 0cm 0cm 0cm},clip]{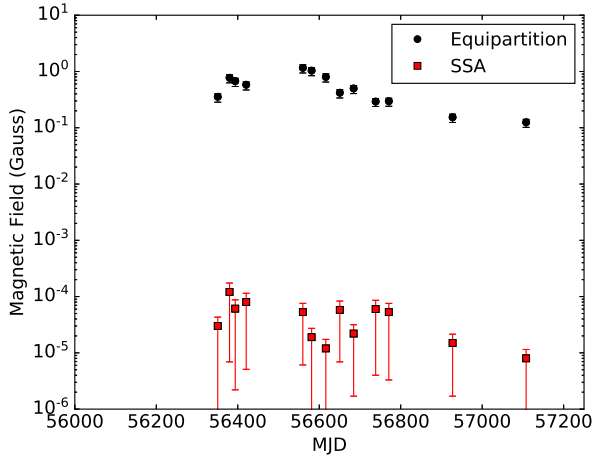} 
\caption{Magnetic fields calculated using the turnover frequency (red squares) and equipartition conditions (black circles)}. 
\label{fig-Bfields}
\end{figure}

To obtain each SSA magnetic field, we performed 10~000 Montecarlo simulations considering variations of the input parameters (i.e., $\Theta_c$, $\delta$, $S_{\nu_c}$ and $\nu_c$) given by their respective uncertainties. Each variable was simulated using a normally distributed random sample, with the mean and standard deviation of the simulated variable being the value and error respectively. A magnetic field value was computed via Equation \ref{eq_B_SSA} for each realization of such simulations. Magnetic fields are given by considering the mean value of the simulations, and the final 1$\sigma$ error is determined by taking the 67 percentiles of the final distribution. This procedure was repeated for each epoch. Table \ref{turnovertable} and Figure \ref{fig-Bfields} summarize the values obtained for $B_{SSA}$ in the different epochs.  Magnetic fields obtained in such way are of the order of 0.1~mG. Although small changes of the magnetic field are seen across different epochs, we find no systematic variation within our uncertainties and the data are in agreement with a roughly constant $B_{SSA}$ over the whole period. 

Alternatively, the magnetic field strength can also be calculated assuming equipartition between the energy of the relativistic particles and the magnetic fields. Following \cite{Kataoka05},
\begin{equation}
\begin{aligned}
B_{eq}&={}123\eta^{2/7}(1+z)^{11/7}\left(\frac{D_L}{100~\mbox{Mpc}}\right)^{-2/7}
\left(\frac{\nu_c}{5~\mbox{GHz}}\right)^{1/7} \\
&\times\left(\frac{S_{\nu_c}}{100~\mbox{mJy}}\right)^{2/7}
\left(\frac{\Theta_c}{\Theta''.3}\right)^{-6/7}
\delta^{-5/7},
\end{aligned}
\end{equation}
where $D_L$ is the luminosity distance and $\eta$ is the ratio of energy density carried by protons and electrons to the energy density of the electrons; i.e., $\eta=1$ for the leptonic jet and $\eta=1836$ for the hadronic jet. Here we assume $\eta\sim100$. Note that there is some debate regarding the relationship between observed sizes and emitting volume in the jet frame, leading to different dependences on the Doppler factor \citep[see e.g.][]{Boettcher12}. Nonetheless, as the dependence is mild, the results will not be significantly affected.

Table \ref{turnovertable} and Figure \ref{fig-Bfields} summarize the values obtained for $B_{eq}$ in the different epochs. Magnetic fields are of the order of $10^2-10^3$~mG. It is clear that the values of  $B_{eq}$ are much larger than these found for $B_{SSA}$, up to a factor $10^4$. This difference cannot be accounted by the large uncertainties or assumptions in the calculations only, and may be providing hints on the physical processes upstream the jet. We will further discuss this possibility  below.

\section{Discussion}
We have studied the time evolution of the innermost radio-structure of 1633+382 during its flux density enhancement period between 2013--2015. High resolution maps show the emergence of two new radio components with an ejection date than can be extrapolated to MJD$=56153\pm30$ and MJD$=56585\pm30$, in good agreement with the epochs of the two largest $\gamma-$ray maxima seen during this study\footnote{We remind here that the a delay of $90\pm40$~days between the $\gamma-$rays and VLBA 43~GHz emission was obtained. Given the uncertainties in both the delay and the ejection dates, this does not qualitatively affect our discussion.}. It is thus natural to associate the flux density enhancement event with the ejection of the new components. Such events have also been observed in several sources, such as 3C~273 \citep{Turler2000,Chidiac16}; 0716+714 \citep{Rani15}; 3C~279 \citep{Lindfors06}; 1156+295 \citep{Ramakrishnan14} among others. The statistical significance of such coincidences has been discussed in \cite[e.g.][]{Jorstad01,LeonTavares11}. In general, when this association is found, it is well in agreement with the prescription given in the shock--in--jet model by \cite{MarscherGear85}.

Interestingly, other local $\gamma-$rays flux maxima are observed in this study (MJD$\sim$56330, MJD$\sim$56650 and MJD$\sim$57040). Weaker than the two major ones, they do not show any indications of a new radio component ejection nor apparent associated core size variability. As discussed in Paper~I, these events did not have a radio counter part, but were tightly correlated with a similar short-lived flux density increase in optical and high energy bands. This supports the discussion previously made suggesting that the nature of these flares must be different, possibly not related with jet activity, and its origin likely located in a different region. Alternatively, the lack of connection between smaller $\gamma-$ray flares and radio flux density could also be due to the $\gamma-$ray and optical emission coming from a small enough portion of the radio-emitting region that the radio flux density is almost unaffected.

This may indicate that (at least) two very different kinds of $\gamma-$ray flares can be traced in this source. First, long-lived (few months) large events that are correlated with radio--jet activity and associated with emerging radio--components; and second, short--lived (few weeks) comparatively weaker events which do not have a radio counterpart but are rather associated with high--energy physics. Further testing is needed to check  whether the long lasting and more energetic $\gamma-$ray flares are always accompanied by radio flares or not. 
 
Different modes of flaring activity have been reported for other sources as well. Two main locations for the origin or the flares have been claimed: near the black hole, ($\ll1$~pc) \citep[see e.g.,][]{Finke10,Tavecchio10,Poutanen10,Foschini11}, or far from the central engine ($\gg1$~pc), where the dusty torus and the jet can be source for the photon fields \citep[see e.g.][]{Lahteenmaki03,Sikora09,LeonTavares11,Tavecchio13}. Even for the same source, events happening at various epochs have been understood as arising from different regions. For example,  the M87 flare in 2005 has been attributed  to the HST-1 complex  \citep{Stawarz06,Cheung07,Harris09} whereas the flares in 2008 and 2012 seem to have originated in the core or jet base \citep{Hada14}. However, there is yet not clear consensus about observational constraints of the $\gamma-$ray flares that could unequivocally pinpoint towards their origin.

In Paper~I we speculated that the emitting region for the large $\gamma-$ray flares was located at around 1~pc from the central engine. The findings in this paper associating the flare with a new VLBI radio component seem to support this scenario. On the other hand, the less powerful, more rapid $\gamma-$ray flares may be associated with smaller regions closer to the central engine, possibly within the broad line region. Alternatively, \cite{Raiteri12} suggested that variations of the viewing angle leading to a change in the Doppler factor may also explain the flux density variation. Since the evolution of radio flux enhancements can be simply explained by radiative processes and does not require additional variation of Doppler factor, our analyses do not favor variations in Doppler as being the possible factor responsible for the observed flux density variability.

\subsection{Source of the Flux Injection}
\label{SourceoftheFluxInjection}
The estimated magnetic fields are of the order of a milli Gauss. The mean values are $\langle B_{SSA}\rangle=0.07$~mG and $\langle B_{eq}\rangle=500$~mG, respectively. The mean equipartition magnetic field is slightly smaller than the value $B_{1}=700\pm300$~mG found in \cite{Algaba12} for 1~pc from the central engine using core shift arguments, and the mean SSA magnetic field is significantly lower. This is reasonable if we consider the regions we are probing in this study seem to be located $\sim40$~pc downstream the jet and consistent with a magnetic field decreasing with distance. Given the uncertainties and the different methodology used, plus the possibility of time variation, we consider that a robust comparison between the magnetic field obtained here and the one in \cite{Algaba12} is not doable.

The estimated magnetic fields seem to be roughly constant within the uncertainties or, if any, slightly decrease after the major $\gamma-$ray flares (i.e, during the period of maximum radio flux density). Even when we consider that the magnetic field has been estimated for different (turnover) frequencies which, by core-shift arguments, would imply they correspond to different regions, analysis in \cite{Algaba12} and  \cite{PaperI} seem to indicate that core shift effects are small in this source and differences in the magnetic field due to this would be smaller than our uncertainties. This indicates that the flux excess that is observed cannot be associated with an increase of the magnetic field strength. Furthermore, compared with the estimated equipartition magnetic field, it appears that  $B_{SSA}\ll B_{eq}$. This further suggests that the magnetic fields may not be playing an important role but rather the source deviates from equipartition in such a way that the balance may be towards an increase of the particle energy density. It thus seems that the observed large and long--lived flares may not be related with strong changes and variability of magnetic fields, but may be associated with a particle dominated region in the base of the jet. The excess of particle energy can be due to injection or acceleration of particles at the base of the jet.

We have to consider several caveats in the discussion. First, calculations for both $B_{SSA}$ and $B_{eq}$  follow certain assumptions, which implies that the resulting values may be over or underestimated. For example, for $B_{eq}$ we assumed $\eta=100$. Depending on the jet composition, $\eta$ may vary, leading to estimated $B_{eq}\sim20$ times smaller or $\sim5$ times larger.  Another caveat is that, as we are not able to resolve the jet components, the spectra that we are using as an input to estimate the magnetic field is the integrated one, and may not trace the actual turnover frequency of the core. As a consequence, the actual SSA turnover frequencies of the core may be higher than the obtained ones, leading to larger $B_{SSA}$ values. Ideally, one should use high resolution VLBI images to de-compose the spectrum into individual components and subtract their effect from the core SED. In practice however, given the high redshift of this source, even the highest resolution VLBI images will be affected by core blending and will still suffer from such caveats. As we described above, the flux density difference between single dish and VLBI images is negligible and, although the exact contribution to the VLBI core from unresolved jet components may be still unclear, it can be estimated to be not significant. Once we take all these caveats into account, it is possible that we may have underestimated $B_{SSA}$ by a factor of few, but not several orders of magnitude. Hence, we consider that these uncertainties do not affect our results qualitatively, and our main conclusion, $B_{SSA} < B_{eq}$, still holds.

If the source were in equipartition, the total energy density would roughly be twice that in the magnetic field, $u=2u_m=2 B^2_{eq}/8\pi\sim0.02$~erg~cm$^{-3}$. If we take the true field (taken to be $B_{SSA}$) to be much smaller than the equipartition one, the true energy density should be larger. Since the energy density $u_p\propto B^{-3/2}$ \citep{CondonRansom16}, then $u=2(B^2_{eq}/8\pi)(B_{SSA}/B_{eq})^{-3/2}\sim1.2\times10^4$~erg~cm$^{-3}$. Following \cite{CelottiFabian93}, $P_{\rm jet}=\pi r_{\rm VLBI}^2 \Gamma^2 \beta c u$, where $r_{\rm VLBI}=(d_{\rm core}/2)[d_{\rm L}/(1+z)^2]$. The jet power is close to $10^{53}$~erg~s$^{-1}$, which is much larger than the synchrotron luminosity of the source, $10^{46}$~erg~s$^{-1}$ based on core flux density measurements at 2, 8, and 15~GHz \citep[e.g. ][]{Lee16a}. We note that these are estimations that may be biased by the caveats described above, matter content, and so on.

Although equipartition is typically considered in the literature, clear deviations have also been found. For example, similar deviation from equipartition was found in other variable compact radio sources by \cite{Jones74} and \cite{Spangler83}, although they did not perform a follow-up during a flaring period. Using a one-zone model, \cite{Inoue96} suggested 3C279 and Mrk~421 to be likely particle dominated. \cite{Kino02} found for the first time robust indications that, for Mrk~421, Mrk~501, and PKS~2155-304, the energy density of relativistic electrons is about an order of magnitude larger than that of magnetic fields. \cite{Homan06} found that, for a large number of sources, the energy in radiating particles exceeded the energy in the magnetic field by a factor of $\sim10^5$ in their maximum brightness state. Similar results were later found by \cite{Migliori11} and \cite{Mankuzhiyil12} for NGC~6251 and Mrk~501, respectively. More recently, \cite{Hayashida15} indicated that modelling the broadband spectral energy distribution in 3C279 during a flare also required the emitting region to be very strongly particle dominated.

For the particular case of 1633+382, \cite{Guijosa96}, found, on the other hand, that $\delta_{eq}/\delta_{IC}=0.39$ in this source. Considering $u_p/u_m=(\delta_{eq}/\delta)^{17/2}\sim0.003$, this suggests that the particle density was four orders of magnitude smaller than the magnetic density, which is the opposite result found here. We note however that these results use data from 1979, which were obtained in a totally different period; and given the variability of the source, we speculate that its core properties may have significantly changed in the last 35 years. Indeed, more recent observations by \cite{Zheng17} found $u_p/u_m\sim20$, which is more consistent with the trend found in our observations.

One final consideration has to be taken into account. During adiabatic expansion losses, where the component size increases with a linear factor $F$, the flux density is expected to decrease at a given frequency as $F^{(4\alpha-2)}$ and the magnetic field as $F^{-2}$, due to flux conservation  \citep[see e.g.][]{ScheuerWilliams68}.  However, based on Figures \ref{fig_components} and \ref{fig-Bfields}, neither the increase of size neither the magnetic field decrease seem a priori clear. We can estimate this in a more quantitative way. If we consider a flux density decrease of a factor of $\sim2$ and a flat spectral index $\alpha=0$, then $F\sim1.4$ and the magnetic field would decrease by a factor of $\sim2$. Whereas we definitively do not see such scaling factor in the core size, this may be due to a combination of observational effects, such as core blending, as discussed above. On the other hand, it is clear that, given the uncertainties of the estimated magnetic fields, we are not sensitive to such changes. Better cadence and well-resolved multi--frequency data to obtain more accurate turnover frequency data and its evolution will be needed.

\section{Conclusions}

In this paper, the second of a series to study the variability of 1633+382, we focus on the radio properties of the source. We have studied structural changes of the source,the miliarc second scale radio morphology,  spectral energy distribution and magnetic fields evolution. By associating these physical properties with the  observed $\gamma-$ray flares, we are able to investigate the origin and physical mechanisms that produce this high energy flux enhancement.

The VLBI data resolved various components moving away from the core. Two of them, C2 and C3, with speeds of $10.2\pm0.8$ and $11.7\pm1.6$~c, have extrapolated ejection epochs MJD$=56520\pm30$ and MJD$=56185\pm30$ respectively, which fall well within the epochs for which the largest $\gamma-$rays were observed. This seems to indicate that the $\gamma-$ray flaring is tightly associated with the ejection of these components. There are no radio structural changes associated with the dimmer $\gamma-$ray flares. The reported flaring activity in the source could be simple explained by radiative processes having a constant Doppler factor.

The turnover frequency shifts towards higher frequencies, from few GHz to few tens of GHz, after the more luminous, long-lived $\gamma-$ray flares occur. The evolution of the flares is in general agreement with the models proposed in \cite{MarscherGear85}. The evolution of the flare in the turnover frequency -- turnover flux density ($S_{\nu_c} - \nu_c$) plane shows an initial complicated pattern for the Compton and synchrotron losses stages due to the overlap of the effects due to two interleaved flares, while the adiabatic losses stage is very clear, with a slope $\epsilon_{adiabatic}=0.6\pm0.1$, which is in agreement  with the model, within the uncertainties.

Estimated magnetic field strength via synchrotron self absorption considerations does not significantly vary over time and of the order of 0.1~mG, smaller by a factor $10^4$ than the magnetic field strength estimated using equipartition arguments. These two findings suggest that the emitting region of the flares is particle dominated.

\acknowledgments
\footnotesize{\emph{Acknowledgements.} We are grateful to all staff members in KVN who helped to operate the array and to correlate the data. The KVN is a facility operated by the Korea Astronomy and Space Science Institute. The KVN operations are supported by KREONET (Korea Research Environment Open NETwork) which is managed and operated by KISTI (Korea Institute of Science and Technology Information). This study makes use of 43 GHz VLBA data from the VLBA-BU Blazar Monitoring Program (VLBA-BU-BLAZAR), funded by NASA through the {\it Fermi} Guest Investigator Program. The VLBA is an instrument of the National Radio Astronomy Observatory. The National Radio Astronomy Observatory is a facility of the National Science Foundation operated by Associated Universities, Inc. This research has made use of data from the OVRO 40-m monitoring program (Richards, J. L. et al. 2011, ApJS, 194, 29) which is supported in part by NASA grants NNX08AW31G, NNX11A043G, and NNX14AQ89G and NSF grants AST-0808050 and AST-1109911. This work used Submillimeter Array data. The Submillimeter Array is a joint project between the Smithsonian Astrophysical Observatory and the Academia Sinica Institute of Astronomy and Astrophysics and is funded by the Smithsonian Institution and the Academia Sinica. G. Zhao is supported by Korea Research Fellowship Program through the National Research Foundation of Korea (NRF) funded by the Ministry of Science, ICT and Future Planning (NRF-2015H1D3A1066561). D.-W. Kim and S. Trippe acknowledge support from the National Research Foundation of Korea (NRF) via grant NRF-2015R1D1A1A01056807. J. C. Algaba and J. Park acknowledge support from the NRF via grant 2014H1A2A1018695. S. S. Lee and S. Kang were supported by the National Research Foundation of Korea (NRF) grant funded by the Korea government (MSIP) (No. NRF-2016R1C1B2006697). This research was supported by an appointment to the NASA Postdoctoral Program at the Goddard Space Flight Center, administered by Universities Space Research Association through a contract with NASA. 
The {\it Fermi}/LAT Collaboration acknowledges the generous support of a number of agencies
and institutes that have supported the {\it Fermi}/LAT Collaboration. These include the National
Aeronautics and Space Administration and the Department of Energy in the United States, the
Commissariat \`a l'Energie Atomique and the Centre National de la Recherche Scientifique / Institut
National de Physique Nucl\'eaire et de Physique des Particules in France, the Agenzia Spaziale
Italiana and the Istituto Nazionale di Fisica Nucleare in Italy, the Ministry of Education,
Culture, Sports, Science and Technology (MEXT), High Energy Accelerator Research Organization
(KEK) and Japan Aerospace Exploration Agency (JAXA) in Japan, and the K.\ A.\ Wallenberg
Foundation, the Swedish Research Council and the Swedish National Space Board in Sweden.
Additional support for science analysis during the operations phase is gratefully acknowledged 
from the Istituto Nazionale di Astrofisica in Italy and the Centre National d'\'Etudes Spatiales 
in France. We are grateful to A. Marscher and S. Jorstad for very useful comments and discussion regarding the magnetic field estimation. We thank the anonymous referee for useful comments and suggestions that helped to improve the manuscript.}



\end{document}